\documentclass[useAMS,usenatbib]{mn2e}

\usepackage{mathptmx}

\usepackage{graphicx}
\usepackage[hypertex]{hyperref}
\usepackage{astrojournals}
\usepackage{amssymb}
\usepackage{comment}
\usepackage{pst-all}
\usepackage{amsmath}
\usepackage{nicefrac}
\usepackage{color}

\def\mnraswidth{8cm}

\newcommand{\unit}[1]{\mathrm{#1}}

\newcommand{\sbnn}{{\sc starburst99}}
\newcommand{\MSolar}{\ensuremath{\textrm{M}_{\odot}}}

\newcommand{\HII}{H\,{\sc ii}}

\newcommand{\Halpha}{\ensuremath{\mathrm{H}{\alpha}}}

\newcommand{\Mspy}{\ensuremath{\,\MSolar\,\unit{yr}^{-1}}}

\newcommand{\modified}{}

\title[Infrared Spectra of Galaxies]{A Semi-Empirical Model of the Infrared Emission from Galaxies}

\author[]{D.~C.~Ford\thanks{Email: dcf21@mrao.cam.ac.uk}, B.~Nikolic and P.~Alexander \\
Astrophysics Group, Cavendish Laboratory, J J Thomson Avenue,
Cambridge CB3 0HE, UK}

\date{Accepted 2008 April 17.  Received 2008 March 19; in original form 2007 July 14}

\pagerange{\pageref{firstpage}--\pageref{lastpage}; } \pubyear{2008}

\begin{document}

\label{firstpage}
\maketitle

\begin{abstract}
We present a semi-empirical model for the infrared emission of dust around
star-forming sites in galaxies. Our approach combines a simple model of
radiative transfer in dust clouds with a state-of-the-art model of the
microscopic optical properties of dust grains pioneered by Draine \& Li. In
combination with the \sbnn\ stellar spectral synthesis package, this framework
is able to produce synthetic spectra for galaxies which extend from the Lyman
limit through to the far-infrared. We use it to probe how model galaxy spectra
depend upon the physical characteristics of their dust grain populations, and
upon the energy sources which heat that dust. We compare the predictions of our
model with the 8- and 24-$\micron$ luminosities of sources in the {\it Spitzer}
First Look Survey, and conclude by using the models to analyse the relative
merits of various colour diagnostics in distinguishing systems out to a
redshift of 2 with ongoing star formation from those with only old stellar
populations.
\end{abstract}


\section{Introduction}
\label{sec:intro}

It is increasingly clear that the infrared spectra of galaxies hold vital clues
concerning galaxy energetics.  Observations of our nearest neighbours tell us
that around 60 per cent of their star formation is obscured by dust at visible
wavelengths \citep{2006A&A...448..525T}, and the cosmic infrared background
(CIB) indicates by its brightness that this is also true of sources at
cosmological redshifts \citep{2001ARA&A..39..249H}.  Furthermore, deep
sub-millimetre surveys
\citep{1997ApJ...490L...5S,1998Natur.394..241H,1998ApJ...507L..21S,1999MNRAS.302..632B}
have revealed a large population of ultra-luminous infrared galaxies (ULIRGs)
at $z \approx 2$ \citep{2002PhR...369..111B} -- clearly very dusty systems,
and, if not {\modified harbouring} active nuclei, very actively star-forming also
\citep{1997ApJ...490L...5S}.  Given the degree of optical extinction in these
systems \citep{2004ApJ...617...64S}, it is apparent that UV/optical studies of
the cosmic star formation history are subject to substantial incompleteness.

The advent of the {\it Spitzer Space Telescope} has allowed a much greater
understanding of the sources which comprise the CIB. Its resolution and
sensitivity has allowed {\modified more than two million infrared
galaxies to be resolved} in $49\,\unit{deg}^2$ of sky by the {\it Spitzer} Wide-Area Infrared
Extragalactic (SWIRE) Survey \citep{2003PASP..115..897L,2004ApJS..154...54L}.
Moreover, {\it Spitzer}'s wavelength coverage, $3.6$--$160\,\micron$,
encompasses three emission regimes in the spectra of normal galaxies, each
yielding information complementary to the others.  In the rest-frame
far-infrared (FIR) -- here taken to extend from around 30 to $300\,\micron$ --
thermal emission from large dust grains dominates. In the rest-frame
mid-infrared (MIR) -- here taken to extend from around 4 to $30\,\micron$ --
emission from transiently-heated dust grains dominates, marked by a series of
broad emission features \citep[see, e.g.,][and references
therein]{2003ARA&A..41..241D}, the catalogue of which has recently been greatly
expanded by {\it Spitzer} \citep{2007ApJ...657..810D,2007ApJ...656..770S}.
These are attributed to polycyclic aromatic hydrocarbon (PAH) molecules, and so
we shall refer to them as `PAH features'\footnote{In the literature, they are
also commonly referred to as `Aromatic Infrared Bands' (AIBs), or,
historically, as `Unidentified Infrared Bands' (UIBs)}.  Finally, in the
rest-frame near-infrared (NIR) -- here taken to refer to
$\lambda\lesssim4\,\micron$ -- stellar emission dominates.

This wealth of available information has motivated many studies which have
sought to provide a framework in which this emission can be interpreted.  Some
of these take an empirical approach, matching unresolved sources to template
spectra derived from a variety of local galaxies \citep{1989MNRAS.238..523R,
2001ApJ...562..179X, 2004MNRAS.351.1290R, 2005AJ....129.1183R}.  These yield
fast diagnostics which are readily applicable to large numbers of sources. But
they provide little information about the physical processes which
fundamentally shape the spectra.

Others have sought to develop semi-empirical models of infrared spectra,
considering the propagation of radiation through dusty media
\citep[e.g.][]{1998ApJ...509..103S, 2000MNRAS.313..734E, 2003PASJ...55..385T,
1995MNRAS.273..649E}. This task can be split into two parts: modelling the
optical properties of individual dust grains, and modelling the large-scale
transport of radiation through some realistic dust geometry.  Some authors
\citep[e.g.][]{2006MNRAS.366..767F, 2006MNRAS.366..923P, 2006MNRAS.370.1454P}
have incorporated a detailed consideration of the radiative transport,
including a treatment not only of absorption, but also of elastic photon
scattering, which becomes the principal source of complexity.  Others
\citep[e.g.][]{2001ApJ...554..778L,2007ApJ...657..810D} have developed highly
sophisticated models of the optical properties of individual dust grains, but
simplify radiative transfer.  Essentially a computational trade-off exists:
studies which include a detailed treatment of radiative transfer involve many
evaluations of dust emissivities in differing environments. This is
prohibitively time consuming with state-of-the-art models of the microscopic
optical properties of individual dust grains and so simpler alternatives must
be sought.

In this paper, we use a state-of-the-art model for the individual dust grains
-- that of \cite{2001ApJ...551..807D} and \cite{2001ApJ...554..778L} -- and
then seek to make a minimal set of simplifications in our treatment of the
radiative transfer problem such that it becomes computationally viable.

In Section~\ref{sec:modelling} we outline the dust geometries which we
consider. In Section~\ref{sec:rad_trans} we go on to present our modelling of
the radiative transfer, and in Section~\ref{sec:radfield} we construct
UV--visible radiation fields appropriate for dust-heating by star-forming
galaxies. In Sections~\ref{sec:dust_pop} and \ref{sec:dust_energetics} we draw
upon models of dust grain populations and their microscopic optical properties
from the literature. In Section~\ref{sec:metallicity} we conclude the
development of our model with a simple framework for modelling the evolution of
the metallicity and mass of gas in passively-evolving star-forming gas clouds.
In Sections~\ref{sec:predictions} and \ref{sec:predthick} we present the basic
predictions of our model. In Section~\ref{sec:validation} we use the model to
predict 8- and 24-\micron\ luminosities for a sample of galaxies and compare to
observations.  Finally, in Section~\ref{sec:colours}, we develop a simple model
for the evolving colours of early- and late-type galaxies.

Wherever required, we assume a flat $\Lambda_\mathrm{CDM}$ cosmology with
$H_0=72\,\unit{km\,s^{-1}\,Mpc^{-1}}$ and $\Omega_\Lambda = 0.7$.

\section{The model geometry}
\label{sec:modelling}

In this paper we consider two basic geometries for the spatial distribution of
dust and the source of illumination.  The first geometry is a shell of dust
grains -- which we shall term a `circumnuclear' grain population -- surrounding a
point-like heating source, representing dust around a star-forming region. The
second is a uniform distribution of dust grains -- which we shall term a
`diffuse' grain population -- spread throughout a diffuse inter-stellar medium
(ISM) within which the radiation field is assumed spatially uniform.  In a
future paper, we shall use them as components of a composite model for
star-forming galaxies.

The geometry adopted for our diffuse populations is the simpler. We assume the
dust-bearing ISM to be optically thin at infrared wavelengths, such that the
re-absorption of dust emission can be neglected. We further assume the
UV--visible heating radiation field within it to be spatially uniform, hence
all grains of any given size and composition have the same emissivity.  Under
these conditions, total dust emission is directly proportional to the number of
grains present, and we therefore scale all quantities per unit volume of ISM.

\begin{figure}
\includegraphics[width=\mnraswidth]{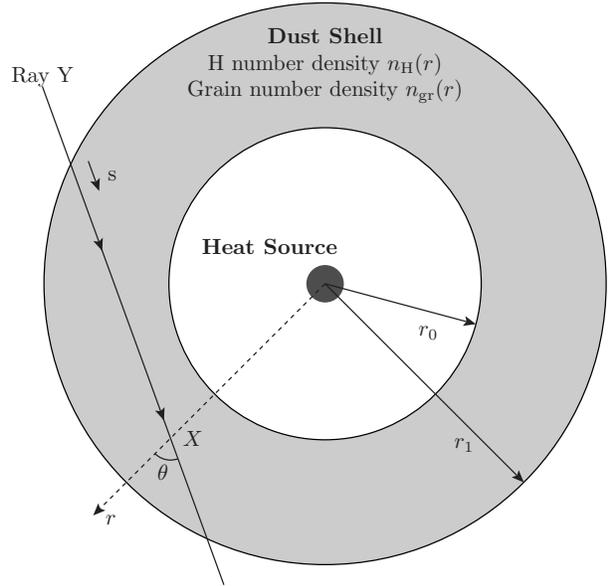}
\caption{Dust in a spherical shell of dust around a single isotropic heating source.}
\label{dust_sphershell}
\end{figure}

The circumnuclear geometry is illustrated in Figure~\ref{dust_sphershell}. A
point-like heating source lies at the centre of a spherical shell of dust, of
inner radius $r_0$ and outer radius $r_1$.  Within the shell, we trace the
density of the medium via the number density of hydrogen nuclei, assumed to be
spherically symmetric and denoted $n_\mathrm{H}(r)$, such that the column
density $N_\mathrm{c}$ of hydrogen nuclei along a line of sight passing through
a dust shell to its nucleus is given by:
\begin{equation}
N_\mathrm{c} =
\int_{r_0}^{r_1} n_\mathrm{H}(r) \, dr .
\end{equation}
Astrophysically, the cavity at $r<r_0$ might correspond to an \HII\ region,
essentially devoid of dust due to the sublimation of grains by energetic
photons.

\section{Radiative Transfer}
\label{sec:rad_trans}

We adopt a highly-simplified treatment of the radiative transfer which we argue
is sufficiently accurate for a wide range of problems.

\subsection{The circumnuclear geometry}
\label{sec:rad_trans_cn}

The evolution of the surface brightness $I_\nu$ along ray $Y$ in
Figure~\ref{dust_sphershell} is governed by the time-independent equation of
radiative transfer:
\begin{equation}
\left.
\frac{\mathrm{d}I_\nu}{\mathrm{d}s}\right|_X
= -C_{\nu,\mathrm{ext}}n_\mathrm{H}(r)I_\nu
+\epsilon_{\nu}(r) n_\mathrm{H}(r)
+\frac{n_\mathrm{H}(r)}{4\pi} \int I_\nu^\prime C_{\nu,\mathrm{sca}}(\theta^\prime) \,\mathrm{d}\Omega^\prime ,
\label{eq:radtrans}
\end{equation}
where the first term on the right-hand side describes the absorption of
radiation by dust, the second dust emission, and the third the scattering of
photons into the ray. The integral in the third term is over
solid angle $\Omega^\prime$ at $X$; $\theta^\prime$ is the scattering angle
between $\mathrm{d}\Omega^\prime$ and the direction of $Y$.

As mentioned above, we use $n_\mathrm{H}(r)$ to parameterise the varying
spatial density of material.  $C_{\nu,\mathrm{abs}}$ and $C_{\nu,\mathrm{sca}}$
are the cross sections to {\modified absorption} and scattering respectively, expressed per
hydrogen atom, and averaged as described below over the compositions and sizes
of particles within the grain population.  $\epsilon_{\nu}(r)$ is the
emissivity of the grain population -- the power emitted per unit frequency into
unit solid angle, normalised in the same way as for the cross sections. $s$
measures distance along the ray.

The extinction cross section, $C_{\nu,\mathrm{ext}}$, is the sum of the
absorption and scattering cross sections:
\begin{equation}
C_{\nu,\mathrm{ext}} = C_{\nu,\mathrm{abs}} + C_{\nu,\mathrm{sca}} .
\end{equation}
The averaging of the quantities above over
grains of varying compositions and sizes is performed as follows:
\begin{eqnarray}
C_{\nu,j} & = & \sum_i \int_a C_{\nu,j}^i(a)
\left( \frac{1}{n_\mathrm{H}} \frac{\mathrm{d}n_\mathrm{gr}^i(a)}{\mathrm{d}a} \right) \,\mathrm{d}a,
\\
\epsilon_{\nu}(r) & = & \sum_i \int_a \epsilon_{\nu}^i(a,r)
\left( \frac{1}{n_\mathrm{H}} \frac{\mathrm{d}n_\mathrm{gr}^i(a)}{\mathrm{d}a} \right) \,\mathrm{d}a,
\label{eq:epsilon_average}
\end{eqnarray}
where $j \in \{ \mathrm{ext},\mathrm{abs},\mathrm{sca} \}$, $i$ denotes a
population of grains of given composition, and $n_\mathrm{gr}^i(a)$ denotes the
spatial number density of grains of each composition with radius smaller than
$a$.

The first simplification that we make is to neglect scattering, such that
$C_{\nu,\mathrm{sca}}=0$. {\modified We discuss the validity of this
assumption in the following section.}
Integrating the remaining terms of
Equation~(\ref{eq:radtrans}) over all rays passing through the point $X$ in
Figure~\ref{dust_sphershell}, we obtain
\citep{BOOK_Chandra1960,1980ApJS...44..403R}:
\begin{equation}
\frac{1}{r^2}\frac{\partial}{\partial r} \left( r^2 H_\nu(r) \right) =
n_\mathrm{H}\left( \epsilon_{\nu}(r) - C_{\nu,\mathrm{abs}} J_\nu(r) \right) ,
\label{eq:zeromoment}
\end{equation}
where:
\begin{equation}
J_\nu(r) = \frac{1}{2} \int_{-1}^{1} I_\nu \mathrm{d}\mu, \quad \mu = \cos \theta,
\end{equation}
and:
\begin{equation}
H_\nu(r) = \frac{1}{2} \int_{-1}^{1} I_\nu \mu \mathrm{d}\mu,
\end{equation}
with $\theta$ denoting the angle made between each ray and the radial
direction, as shown in Figure~\ref{dust_sphershell}.

Geometrically, $J_\nu(r)$ may be visualised as the average surface brightness
along all rays passing through $X$, averaged over $4\pi$ steradians.
$H_\nu(r)$ may similarly be visualised as the average projected onto the radial
direction.

\cite{1980ApJS...44..403R} introduced what has become a widely-used
decomposition of the surface brightness $I_\nu$ along each ray into three
components, depending upon where photons last interacted with matter:
\begin{equation}
I_\nu = I_\nu^{(1)} + I_\nu^{(2)} + I_\nu^{(3)} ,
\label{eq:rr_decomp}
\end{equation}
where $I_\nu^{(1)}$ is radiation from the central heat source which has not
been absorbed by dust, $I_\nu^{(2)}$ is radiation emitted by dust and
$I_\nu^{(3)}$ is scattered radiation -- which we have already neglected. While
we do not use this in our mathematical treatment of
Equation~(\ref{eq:radtrans}), it is useful in our discussion presently.

The relationship between $J_\nu(r)$ and $H_\nu(r)$ encodes the angular
distribution of the radiation flux passing through $X$. In the limiting case of
a radiation field propagating exclusively in the radial direction,
$H_\nu(r)=J_\nu(r)$. In the opposite limit of an isotropic radiation field,
$H_\nu(r)=\nicefrac{1}{2}J_\nu(r)$. Given a point heat source, as in
Figure~\ref{dust_sphershell}, the former limit is applicable to the component
$I_\nu^{(1)}$; the radiation field emanating from the heat source is purely
radial. For $I_\nu^{(2)}$, however, $H_\nu(r)<J_\nu(r)$.

Our second, and final, assumption, is that $J_\nu(r)=H_\nu(r)$ in
Equation~(\ref{eq:zeromoment}). For UV--visible wavelengths, this assumption
holds because the heating radiation field is expected to dominate over dust
emission at these wavelengths, and so $I_\nu \approx I_\nu^{(1)}$. In the
infrared, where $I_\nu^{(2)}$ is expected to dominate $I_\nu$, $J_\nu(r)$ is
{\modified under-predicted}, but the assumption continues to hold if
$\epsilon_{\nu}(r) \gg C_{\nu,\mathrm{abs}} J_\nu(r)$, that is to say, if the
dust emission from the shell is not appreciably re-absorbed.  Geometrically,
this assumption is equivalent to assuming that dust emission is beamed along
the outward radial direction.

Finally, we note that the net outward flux $F_\nu(r)$ of radiation through the
sphere of constant radius passing through $X$ is related to $H_\nu(r)$ via:
\begin{equation}
F_\nu(r) = 4\pi H_\nu(r).
\end{equation}
Equation~(\ref{eq:zeromoment}) can thus be re-written:
\begin{equation}
\frac{1}{r^2}\frac{\partial}{\partial r} \left( r^2 F_\nu(r) \right) =
n_\mathrm{H}(r) \left(
4 \pi \epsilon_\nu(r) - C_{\nu,\mathrm{abs}} F_\nu(r)
\right) .
\label{eq:radtrans_diffeq}
\end{equation}
This equation is integrated numerically from the inner to the outer radius.

{\modified
\subsection{Assumptions made in the circumnuclear geometry}

The assumptions made in the previous section -- i.e.\ the neglect of scattering
and the radial beaming of dust emission -- will have negligible effect upon the
predictions of our model for column densities of dust which are optically thin
at all wavelengths, i.e.\ for
$N_\mathrm{c}\lesssim10^{23}\,\unit{H}\,\unit{m}^{-2}$.  For column densities
of dust which are optically thick in the UV, but not in the infrared, our
neglect of scattering will lead us to under-estimate the path lengths of
UV/optical photons through the dust shell by a factor of 1--2, and so to
under-estimate the absorption of UV/optical radiation by a similar factor.
Since this effect is essentially the same as that of reducing the column
density of dust, the effect when using these models to fit the spectral shape
of real sources will be that we will over-estimate the dust masses of these
objects.

Our assumption that dust emission is beamed radially outwards only begins to
fail for dust shells with higher column densities still, when they become
optically thick even at infrared wavelengths. As the dust emission is assumed
to take the shortest path out of the dust shell, we will under-estimate its
re-absorption in these optically thick cases.  In practice,
this effect becomes significant for dust column densities $\gtrsim
10^{26}\,\unit{H}\,\unit{m}^{-2}$, as will be shown in Figure~\ref{fig:gal2d2}.

\cite{1980ApJS...44..403R} studied the effects of a similar set of assumptions
in their calculation of radiative transport in hot-centred star-forming clouds,
and for the range of systems they consider, they report errors of around 10~per
cent.

In addition to the two assumptions just discussed, it is also apparent that the
adopted geometry is simplistic, having only one single heat source. We note,
however, that this geometry is observationally indistinguishable from an
ensemble of $N$ smaller circumnuclear geometries, each heated by its own
central heat source with luminosity scaled by a factor $1/N$ with respect to
the single shell, and each containing a dust shell with inner and outer radii
scaled by factor $1/\sqrt{N}$ and dust density distribution $n'(r)$ scaled
according to:
\begin{equation}
n'(r) = \sqrt{N} n(r\sqrt{N})
\end{equation}
where $n(r)$ is the density distribution of the single shell.

The column density of dust around each heat source in this latter ensemble is
the same as that in the former single circumnuclear shell; the radiation field
incident upon grains on the inner edge of each dust shell is the same; the
total mass of dust in the two cases is the same; and to a remote observer, the
total solid angle subtended by the dust in the two cases is the same. In
summary, although our circumnuclear model is nominally of dust around a single
heat source, it is also a good model of systems where that luminosity
production is distributed between several discrete sources.

}

\subsection{The diffuse geometry}
\label{sec:rad_trans_dif}

Our treatment of radiative transfer in diffuse dust grain populations is
simpler than the above.  Instead of having a heat source of luminosity $L_\nu$,
we have an interstellar radiation field (ISRF), whose energy density
$E_\nu^{(1)}$ we normalise to $\chi\chi_0$, where $\chi_0$ is that of the solar
neighbourhood interstellar radiation field less the cosmic microwave background
(CMB)\footnote{We neglect the CMB in this normalisation because, in contrast to
the starlight component of the ISRF, it would make no sense to enhance it by a
factor $\chi$. It should be noted that the CMB is also absent from all models
presented in this paper.}, and $\chi$ is a free parameter. We adopt
$\chi_0=7.46\times 10^{-14}\,\unit{J}\,\unit{m}^{-3}$, derived from integration
of the ISRF of \cite{1983A&A...128..212M} {\modified and} \cite{1982A&A...105..372M}.

The total luminosity emerging from the model can then be written:
{\modified
\begin{equation}
L_\nu = 4 \pi \epsilon_\nu n_\mathrm{H} V  +  \frac{E_\nu^{(1)} c A}{4}
\label{eq:diff_lum}
\end{equation}
}
where $V$ is the volume of the dust-bearing ISM, and $A$ its surface area
through which the interstellar radiation field leaves the galaxy.

\section{The heating radiation field}
\label{sec:radfield}

In this paper, we consider models for the heating radiation field due to star
formation.  We generate these using Version~5.1 of the \sbnn\ stellar spectral
synthesis package \citep{1999ApJS..123....3L, 2005ApJ...621..695V}. This
package offers the facility to model stellar populations with two classes of
star formation histories (SFHs).  The first is ongoing star formation,
proceeding at constant star-formation rate (SFR) $\Psi$, which began at some
time $t$ previously; we hereafter term these models `continuous' SFHs. The
second is a delta-function SFH, representing an instantaneous burst of star
formation at some time $t$ previously, of total mass $m$; we term this the
`instantaneous' SFH. The stellar populations modelled for the instantaneous
SFHs may be referred to as single stellar populations (SSPs), which is to say
that all of the stars within them are coeval.

\sbnn\ models both stellar emission and also nebular continuum. For the former
it uses the stellar evolution tracks of the Padova group
\citep{1994A&AS..105...29F}, with the addition of tracks for thermally pulsing
asymptotic giant branch {\modified (TP-AGB)} stars to improve the accuracy of
the modelling of {\modified low and intermediate mass stars}.\footnote{It
should be noted that this is a significant departure from previous versions of
\sbnn, which used the stellar evolution tracks of the Geneva group, and did not
model low mass stars, introducing serious errors in the modelling of old
stellar populations.} For the nebular continuum, the emission coefficients of
\cite{1980PASP...92..596F} are used.  This is the source of a problem in
\sbnn\footnote{See notice by Hunt, October 30, 2006, in the {\it Knowledge
Base} of the \sbnn\ website.}: the data of \cite{1980PASP...92..596F} do not
extend beyond $4.5\,\micron$, however they are extrapolated to $160\,\micron$,
yielding large, unphysical, infrared luminosities. For $\lambda\geq4\,\micron$,
we use a more physical extrapolation, taking $L_\nu \propto \nu^{-0.1}$.

In addition to the continuous and instantaneous SFHs modelled by \sbnn, we have
also considered arbitrary SFHs, modelled by convolving the luminosity,
$L_\mathrm{SSP}(t)$, of a $10^6\,\MSolar$ single stellar population of age $t$
with our star formation history:
\begin{equation}
L_\nu (t) = \int_0^{t} \left( \frac{\Psi(t')}{10^6\,\unit{M_\odot}} \right) L_\mathrm{SSP}(t-t') \,\mathrm{d}t' .
\label{eq:folding_scheme}
\end{equation}

\section{The dust model}
\label{sec:dust_pop}

Whilst there exist fairly tight observational constraints on the composition
and size distribution of dust grains in the Milky Way, relatively little is
known about those in other galaxies \citep{2003ARA&A..41..241D}.  In this
paper, we therefore base our dust grain population upon that inferred from
observation of our own galaxy.

Following {\modified \citet[][hereafter, LD01]{2001ApJ...554..778L}}, we
consider binary populations of `carbonaceous' and silicate grains. The former
sub-population includes both PAH molecules and larger graphitic grains; the
optical properties of these grains exhibit a smooth transition with grain
radius, centred around a radius of $a_{\xi}$. The absorption cross sections of
carbonaceous grains of radius $a$ is taken to be:
\begin{equation}
C^\mathrm{car}_{\nu,\mathrm{abs}}(a) = \xi_\mathrm{PAH}(a) C^\mathrm{PAH}_{\nu,\mathrm{abs}}(a) +
[1-\xi_\mathrm{PAH}(a)]C^\mathrm{gra}_{\nu,\mathrm{abs}}(a) ,
\label{eq:axi_transition}
\end{equation}
where the weighting parameter $\xi_\mathrm{PAH}(a)$ is given by:
\begin{equation}
\xi_\mathrm{PAH}(a) = \left( 1 - q_\mathrm{gra} \right) \times \min
\left[ 1, (a_{\xi}/a)^3 \right] ,
\label{eq:xipah}
\end{equation}
and $q_\mathrm{gra} = 0.01$ sets even the smallest PAH molecules to exhibit 1
per cent of the continuum absorption of graphitic grains. The transition
radius, $a_{\xi}$, is set by default to $50\,\mathrm{\AA}$.

For the size-distribution of grains in each of these populations, we follow the
parametric forms used by \cite{2001ApJ...548..296W}; for the carbonaceous
grains, we use:

\begin{equation}
\label{eq_carbsizedist}
\frac{1}{n_\mathrm{H}}
\left(\frac{Z_\odot}{Z}\right)
\left(\frac{\mathrm{d}n_\mathrm{gr}^\mathrm{car}}{\mathrm{d}a}\right)
 =  D(a) + \frac{C_\mathrm{g}}{a}\left(\frac{a}{a_\mathrm{t,g}}\right)^{\alpha_\mathrm{g}}
F(a , \beta_\mathrm{g} , a_\mathrm{t,g})
\end{equation}
\begin{displaymath}
\times \left\{
\begin{array}{llllll}
1, & 3.5\,\mathrm{\AA} & < & a & < & a_\mathrm{t,g} \\
\exp \left\{-\left[\left(a - a_\mathrm{t,g}\right) / a_\mathrm{c,g}\right]^3\right\},
& a_\mathrm{t,g} & < & a & & \\
\end{array}
\right.
\end{displaymath}

\noindent where $D(a)$ represents two log-normal peaks:

\begin{equation}
D(a) = \sum_{i=1}^2 \frac{B_i}{a} \exp \left\{ -\frac{1}{2} \left[
\frac{\ln\left(a/a_{0,i}\right)}{\sigma}
\right]\right\} ,
\end{equation}

{\modified \noindent which were introduced by LD01 to reproduce the mid-infrared luminosities observed by ISO, the {\it Infrared Telescope in Space} (IRTS) and by IRAS at $60\,\micron$.
}
The term $F(a , \beta_\mathrm{g} , a_\mathrm{t,g})$ provides curvature:

\begin{equation}
F(a , \beta_\mathrm{g} , a_\mathrm{t}) =
\left\{
\begin{array}{ll}
1+\beta a / a_\mathrm{t}        & \beta \geq 0 \\
(1-\beta a / a_\mathrm{t})^{-1} & \beta < 0 ,
\end{array}
\right.
\end{equation}

\noindent $Z$ is the mass ratio of metals, $Z_\odot=0.02$ is the solar mass
ratio of metals, and all other symbols are as defined in
\cite{2001ApJ...548..296W}. The normalisation constants $B_i$ are given by:

\begin{equation}
B_i = \frac{3}{(2\pi)^{\nicefrac{3}{2}}}
\frac{\exp\left(-4.5 \sigma^2\right)}{\rho a_{0,i}^3 \sigma}
\frac{b_{\mathrm{C},i} m_\mathrm{C}}
{1 + \mathrm{erf}\left( \frac{3\sigma}{\sqrt{2}} + \frac{\ln\left(a_{0,i} / 3.5\,\unit{\AA}\right) }{\sqrt{2}\sigma}\right)}
\end{equation}

\noindent where $m_\mathrm{C}=1.99\times 10^{-26}\,\unit{kg}$ is the mass of a
carbon atom, $\rho=2.24\times 10^3\,\unit{kg}\,\unit{m}^{-3}$ is the density of
graphite, $a_{0,1}=3.5\,\unit{\AA}$ and $a_{0,2}=30\,\unit{\AA}$ are the
wavelengths of the centres of the two log-normal peaks, and $\sigma=0.4$.

For the silicate grains, we use:

\begin{equation}
\frac{1}{n_\mathrm{H}}
\left(\frac{Z_\odot}{Z}\right)
\left(\frac{\mathrm{d}n_\mathrm{gr}^\mathrm{sil}}{\mathrm{d}a}\right)
 = \frac{C_\mathrm{s}}{a}\left(\frac{a}{a_\mathrm{t,s}}\right)^{\alpha_\mathrm{s}}
F(\alpha ; \beta_\mathrm{s} , a_\mathrm{t,s})
\end{equation}
\begin{displaymath}
\times \left\{
\begin{array}{llllll}
1, & 3.5\,\mathrm{\AA} & < & a & < & a_\mathrm{t,s} \\
\exp \left\{-\left[\left(a - a_\mathrm{t,s}\right) / a_\mathrm{c,s}\right]^3\right\},
& a_\mathrm{t,s} & < & a & & \\
\end{array}
\right.
\end{displaymath}

By default, we set the parameters of these size distributions to those
preferred by \citet{2001ApJ...548..296W} for $R_\mathrm{V} = 3.1$ Milky Way
sight lines, as given in Table~\ref{tab:sizedist_params}. {\modified The effect
of using instead the size distributions preferred by those authors for
$R_\mathrm{V} = 4.0$ and $R_\mathrm{V} = 5.5$ Milky Way sight lines will be
discussed in Section~\ref{sec:predictions}.}

\begin{table}
\begin{center}
\begin{tabular}{p{3.5cm}p{3.5cm}}
\hline
{\bf Parameter} & {\bf Value} \\
\hline
$\alpha_\mathrm{g}$ & $-1.54$  \\
$\beta_\mathrm{g}$  & $-0.165$ \\
$a_{\mathrm{t,g}}$  & $\phantom{-}0.0107\,\micron$ \\
$a_{\mathrm{c,g}}$  & $\phantom{-}0.428\,\micron$ \\
$C_\mathrm{g}$      & $\phantom{-}9.99 \times 10^{-12}$ \\
$\alpha_\mathrm{s}$ & $-2.21$ \\
$\beta_\mathrm{s}$  & $\phantom{-}0.300$ \\
$a_{\mathrm{t,s}}$  & $\phantom{-}0.164\,\micron$ \\
$C_\mathrm{s}$      & $\phantom{-}1.00 \times 10^{-13}$ \\
$b_{\mathrm{C},1}$  & $\phantom{-}0.75b_\mathrm{C}$ \\
$b_{\mathrm{C},2}$  & $\phantom{-}0.25b_\mathrm{C}$ \\
$b_\mathrm{C}$      & $\phantom{-}6.0\times 10^{-5}$ \\
\hline
\end{tabular}
\end{center}
\caption{The parameters of the default size distribution which we adopt -- the \citet{2001ApJ...548..296W} preferred distribution for $R_\mathrm{V} = 3.1$ Milky Way sight lines.}
\label{tab:sizedist_params}
\end{table}

The appearance of $Z$ in Equation~(\ref{eq_carbsizedist}) is a departure from
\cite{2001ApJ...548..296W}, who only considered Galactic environments. Thus,
their size distributions are normalised to a dust-to-gas-mass ratio appropriate
for solar metallicity environments. As the variation in this ratio with $Z$ is
quite poorly understood, we make the assumption that it is linearly
proportional to $Z$, which is implicit in our renormalisation above.

\label{sizedist}
\begin{figure}
  \includegraphics[width=\mnraswidth]{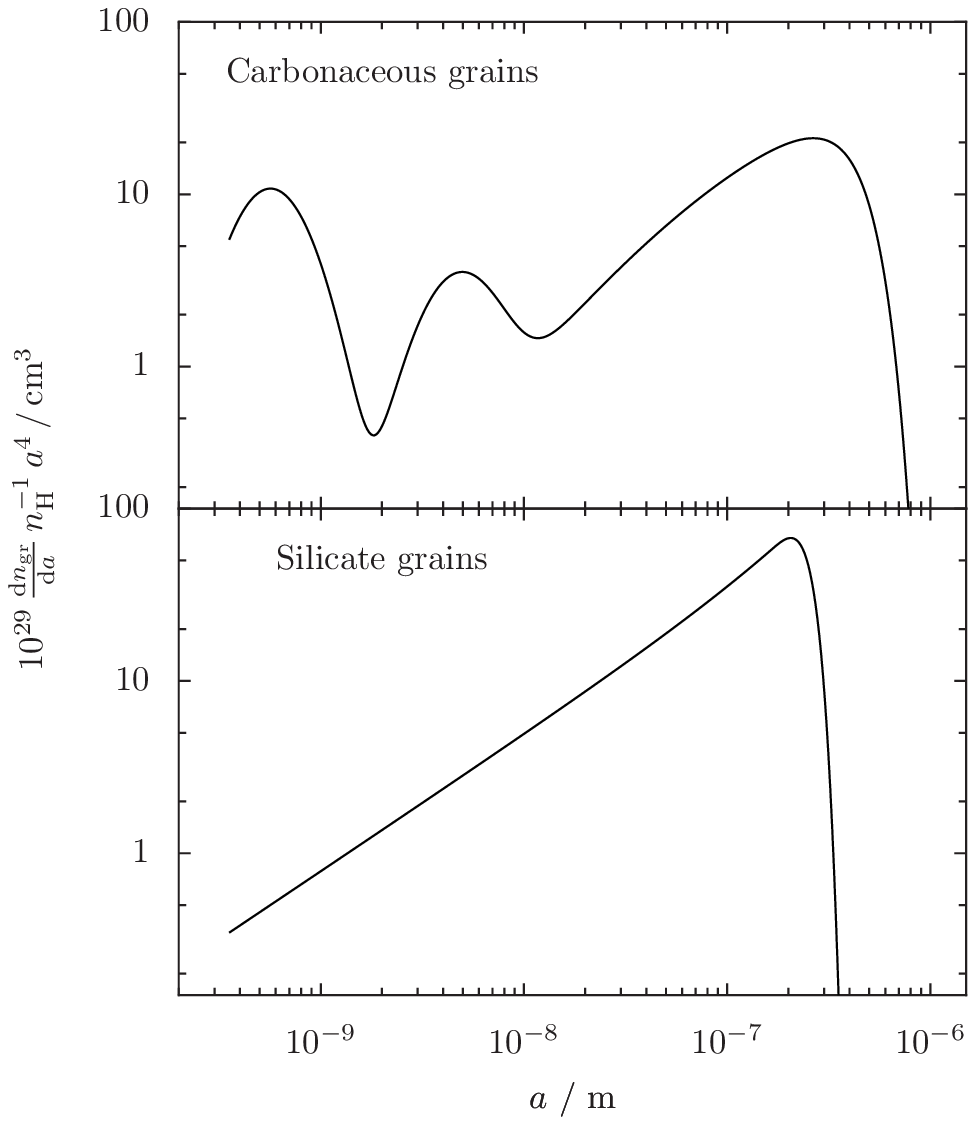}
  \caption{The adopted size distributions for silicate (top) and
  carbonaceous (bottom) grains, assuming a solar metallicity environment.
  Unit areas under each distribution represent unit masses of grain material.}
  \label{fig:sizedist}
\end{figure}

\label{sec:dust_int}

These distributions are plotted in Figure~\ref{fig:sizedist}. To calculate the
absorption cross sections of silicate and graphitic grains, we follow
{\modified LD01} and use dielectric functions for these species
\citep{1984ApJ...285...89D} and Mie theory \citep[see,
e.g.,][]{BOOK_BohrenHuffman} to estimate the absorption cross sections of
spherical particles of radius $a$.  The treatment of graphitic grains is
slightly complicated by the anisotropy of graphite's dielectric function.  We
follow LD01 in calculating an averaged absorption cross section using the
`\nicefrac{1}{3}-\nicefrac{2}{3} approximation' \citep{1993ApJ...414..632D}.

\begin{figure*}
  \includegraphics[width=15.5cm]{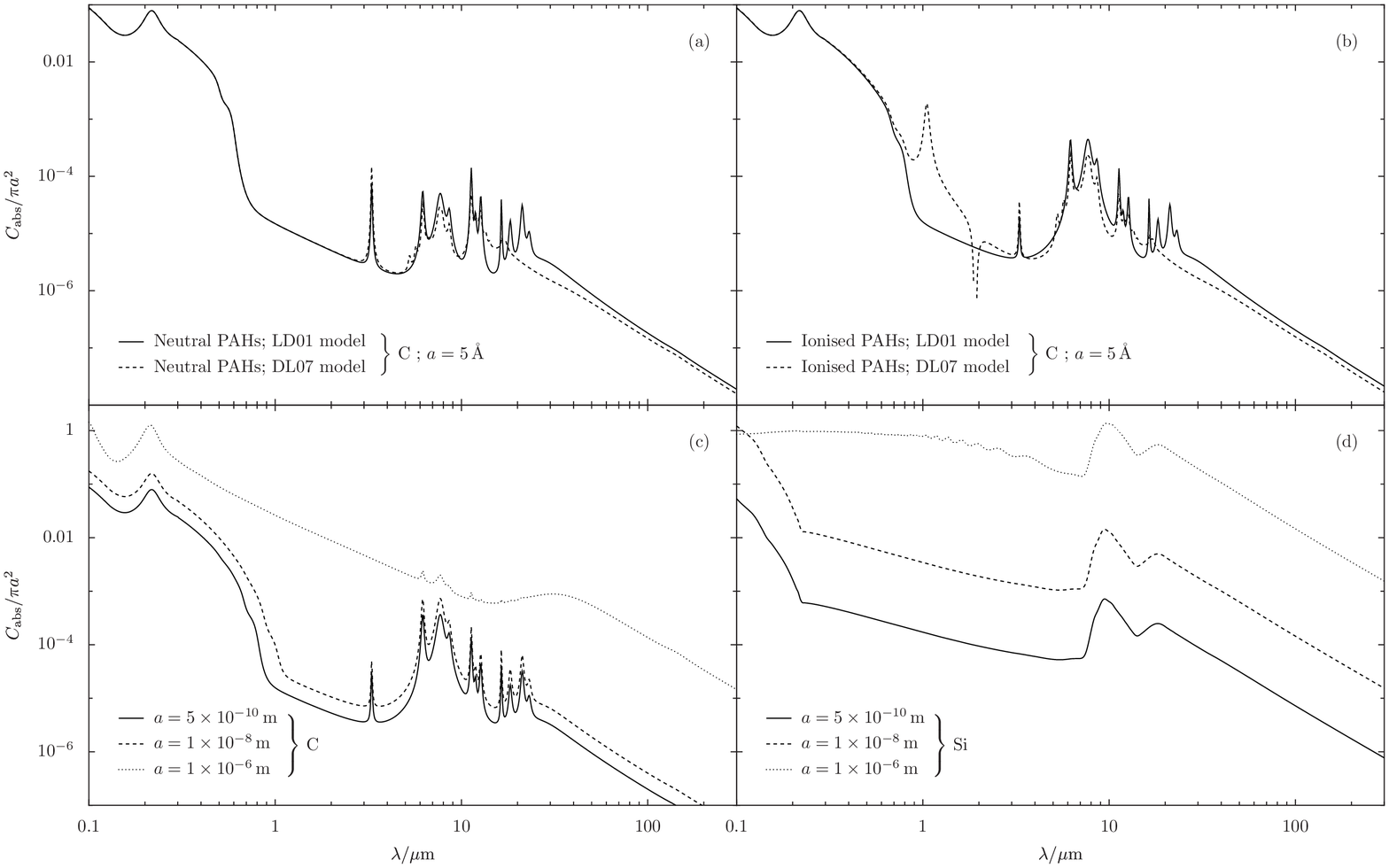}
  \caption{The adopted absorption cross sections. Panels~(a) and (b) show those for
     neutral and ionised PAH grains respectively, as given by LD01 and DL07 for 
     grains of radius $5\,\unit{\AA}$.  A discussion of the near-infrared feature
     introduced into the cross sections of ionised PAHs by DL07 at 
     $1.05\,\micron$, and the negative feature at $1.905\,\micron$, can be found in
	  \citet{2005ApJ...629.1183M}. Panels~(c) and (d) show those for
     carbonaceous and silicate grains respectively for a variety of grain radii $a$,
     assuming a PAH ionisation fraction $f=0.8$. Each trace is normalised with
     respect to the classical grain cross section of $\pi a^2$.
     }
  \label{fig:xsec_vs_CSi}
\end{figure*}

For the PAH molecules, LD01 give algebraic fits to terrestrial laboratory
measurements of $C^\mathrm{PAH}_{\nu,\mathrm{abs}}(a)$ for neutral and ionised
samples. \citet[][hereafter, DL07]{2007ApJ...657..810D} revise these in the
light of new near-infrared data \citep{2005ApJ...629.1188M}, and in order to
fit the high-fidelity spectra of nearby star-forming galaxies observed by the
{\it Spitzer} Infrared Nearby Galaxies Survey
\citep[SINGS;][]{2003PASP..115..928K} project.  We implement the cross sections
given by both LD01 and DL07, which are shown in
Figures~\ref{fig:xsec_vs_CSi}(a) and \ref{fig:xsec_vs_CSi}(b) for neutral and
ionised grains respectively, both of radius $5\,\unit{\AA}$.  In the remainder
of this paper, we use the DL07 cross sections throughout, except in
Section~\ref{sec:validation}.  In both cases, we average the cross sections of
neutral and ionised PAH molecules with a weighting parameter $f$ describing the
ionisation fraction. We take this to have a default value of 80 per cent,
matching that which \cite{2001ApJ...551..807D} find in their model fits to
Galactic {\modified photo-dissociation} regions.  The resulting absorption cross section
$C_{\nu,\mathrm{abs}}^\mathrm{car}(a)$ is shown for a range of grain radii in
Figure~\ref{fig:xsec_vs_CSi}(c), and the absorption cross section
$C_{\nu,\mathrm{abs}}^\mathrm{sil}(a)$ of the silicate grain population
Figure~\ref{fig:xsec_vs_CSi}(d).

\section{Modelling the emissivity of the dust}
\label{sec:dust_energetics}

In this section, we outline how we model the emissivities $\epsilon_\nu^i(a,r)$
of dust grains as a function of the energy density $E_\nu(r)$ of radiation to
which they are subjected.

\subsection{Transiently-heated grains}
\label{sec:transient_heat}

Modelling emission from transiently-heated grains requires the calculation of
the time-averaged probability distributions $P(E)$ for their internal energies.
Exact treatment of this problem would require knowledge of all of their
vibrational energy levels and transition probabilities.  Our simplified
analysis follows \cite{2001ApJ...551..807D}.  We use Debye models for the
normal modes of the C/Si skeletons of PAH and silicate particles, with Debye
temperatures as used by \cite{2001ApJ...551..807D}. We use Einstein models for
the stretching, in-plane bending, and out-of-plane bending modes of the
peripheral C--H bonds of PAH molecules: we assume the modes of each bond to be
quantum harmonic oscillators with the same fundamental frequencies, as given in
\cite{2001ApJ...551..807D}.

To reduce the resulting mode spectra to a computationally tractable number of energy
states, we follow \cite{1989ApJ...345..230G} and \cite{2001ApJ...551..807D} in
dividing them into $N_\mathrm{bin}$ bins (for our choice of bins, see
Appendix~A), with mean energies $\boldsymbol{U}_i$, widths
$\Delta\boldsymbol{U}_i$, and time-averaged occupation probabilities
$\boldsymbol{P}_i$. We denote as $\mathbf{T}_{ji}$ the transition rate between
bins $i$ and $j$. The time evolution of
$\boldsymbol{P}_i$ is then given by:
\begin{equation}
\frac{\mathrm{d}\boldsymbol{P}_i}{\mathrm{d}t} =
\sum_{j\neq i} \mathbf{T}_{ij} \boldsymbol{P}_j -
\sum_{j\neq i} \mathbf{T}_{ji} \boldsymbol{P}_i   .
\label{eq:P_timeevol}
\end{equation}
The time-averaged steady-state probability distribution which we seek is that
to which the above converges over time, and for which
$\mathrm{d}\boldsymbol{P}_i/\mathrm{d}t=0$.

The elements of $\mathbf{T}_{ji}$ with $j>i$ describe the upward transitions of
grains that result from photon absorption; we model these using
equations~(15--25) of \cite{2001ApJ...551..807D}; in Appendix~B we {\modified
reproduce these relations and} describe a numerical optimisation that we use in
their calculation. The elements with $j<i$ describe the radiative cooling of
grains; here we use the `thermal continuous' approximation \citep[equation~41
of][{\modified reproduced here as
Equation~\ref{t_downward}}]{2001ApJ...551..807D}, which models the cooling of
grains as a continuous process, where each state $i$ only makes downward
transitions to the adjacent state $i-1$. This allows much faster solution of
Equation~\ref{eq:P_timeevol} to find $\boldsymbol{P}_i$.

The diagonal terms are chosen {\modified \citep{2001ApJ...551..807D}} so that:
\begin{equation}
\mathbf{T}_{ii} = - \sum_{j\neq i} \mathbf{T}_{ji} ,
\end{equation}
hence Equation~(\ref{eq:P_timeevol}) can be re-written in the form:
\begin{equation}
\frac{\mathrm{d}\boldsymbol{P}_i}{\mathrm{d}t} =
\sum_{j=0}^{N_\mathrm{bin}} \mathbf{T}_{ij} \boldsymbol{P}_j = 0.
\label{eq:P_timeevol2}
\end{equation}
These equations are solved, subject to the additional normalisation constraint:
\begin{equation}
\sum_i \boldsymbol{P}_i=1 ,
\end{equation}
using the method of \cite{1989ApJ...345..230G}.  Given the vector
$\boldsymbol{P}_i$, we calculate the time-averaged emissivity of each grain
using the thermal approximation, under which $\epsilon_\nu^i(a,r)$ can be
calculated using equation~(56) of \cite{2001ApJ...551..807D}:
{\modified
\begin{equation}
\epsilon_\nu = \frac{2h\nu^3}{c^2} \left[
\sum_i \frac{\boldsymbol{P}_i}{\exp(h\nu/k\boldsymbol{\theta}_i)-1}\right],
\end{equation}
where $\boldsymbol{\theta}_i$ is the characteristic temperature of bin $i$, as
defined in \citet{2001ApJ...551..807D}, the sum is over all bins $i$ whose
central energies are greater than $h\nu$, and we have neglected the factor
$(1+\lambda^3u_E / 8\pi)$ shown by those authors; this represents stimulated
emission and may straightforwardly be shown to be negligible in all of the
models presented in this paper.
}

\subsection{Large grains}

For sufficiently large grains, the approach outlined above becomes inefficient.
Their internal energies become much larger than the energies of the photons
they absorb, and so their temperature fluctuations are not significant. Their
internal energy probability distributions $P(E)$ tend towards delta functions
\citep{2001ApJ...554..778L}.  In this limit, we can model the energetics and
emission of these grains by numerically solving the equation of radiative
balance to find their equilibrium temperatures:
\begin{equation}
\int_0^\infty      C_{\nu,\mathrm{abs}}(a) c u_\nu  \, \mathrm{d}\nu =
\int_0^\infty 4\pi C_{\nu,\mathrm{abs}}(a) B_\nu(T) \, \mathrm{d}\nu ,
\label{eq:rad_balance}
\end{equation}
\noindent where $c$ is the speed of light, and $B_\nu(T)$ the Planck function
at temperature $T$. \cite{2001ApJ...554..778L} show that when grains are bathed
in the local ISRF of the Solar Neighbourhood, this continuum approximation is
valid for grain radii greater than $250\,\unit{\AA}$. We adopt this same
transition radius for the range of models considered here.

{\modified
The emissivities of these grains are calculated by assuming them to radiate as
modified blackbodies with a single characteristic temperature:
\begin{equation}
\epsilon_\nu = C_{\nu,\mathrm{abs}} B_\nu(T) .
\end{equation}
}

\section{Evolution of gas mass and metallicity}
\label{sec:metallicity}

To follow the evolution of a star-forming system we use a simple closed box
model. We denote the total stellar mass-loss rate (due to stellar winds and
supernovae) as $\Sigma(t)$, with metallicity $Z_\mathrm{outflow}$. The total
mass of gas and dust in a cloud is $m_\mathrm{c}(t)$, of which metals comprise
a mass $m_\mathrm{z}(t)$, such that $Z(t)=m_\mathrm{z}(t) / m_\mathrm{c}(t)$.
Thus we may write:
\begin{equation}
\frac{\mathrm{d}m_\mathrm{c}}{\mathrm{d}t} = - \Psi(t) + \Sigma(t) .
\end{equation}

We consider two models for the mixing of metal-enriched material in the cloud.
In the simpler, we assume a time-invariant metallicity, such that
$m_\mathrm{z}(t) = Z m_\mathrm{cloud}(t)$. We call this model the `constant
$Z$' mixing model.  In the second, we assume perfect mixing, such that the
metallicity of the material which goes into forming new stars is representative
of that of the whole cloud, in which case:
\begin{equation}
\frac{\mathrm{d}m_\mathrm{z}}{\mathrm{d}t} = - Z(t)\Psi(t) + Z_\mathrm{outflow}(t)\Sigma(t) .
\end{equation}
We call this the `perfect mixing' model.

To calculate $\Sigma(t)$ and $Z_\mathrm{outflow}(t)$ we used the same \sbnn\
models as described in Section~\ref{sec:radfield}.  The evolving metallicity of
the gas forming into new stars could not be treated smoothly, as the Padova
stellar-evolution tracks used by \sbnn\ are only available for five stellar
metallicities: $Z=0.0004$, 0.004, 0.008, 0.02 and 0.05. Instead, we modelled
each SSP within our calculation of the SFH using the track whose metallicity
was logarithmically closest to that desired.

\begin{figure}
  \includegraphics[width=\mnraswidth]{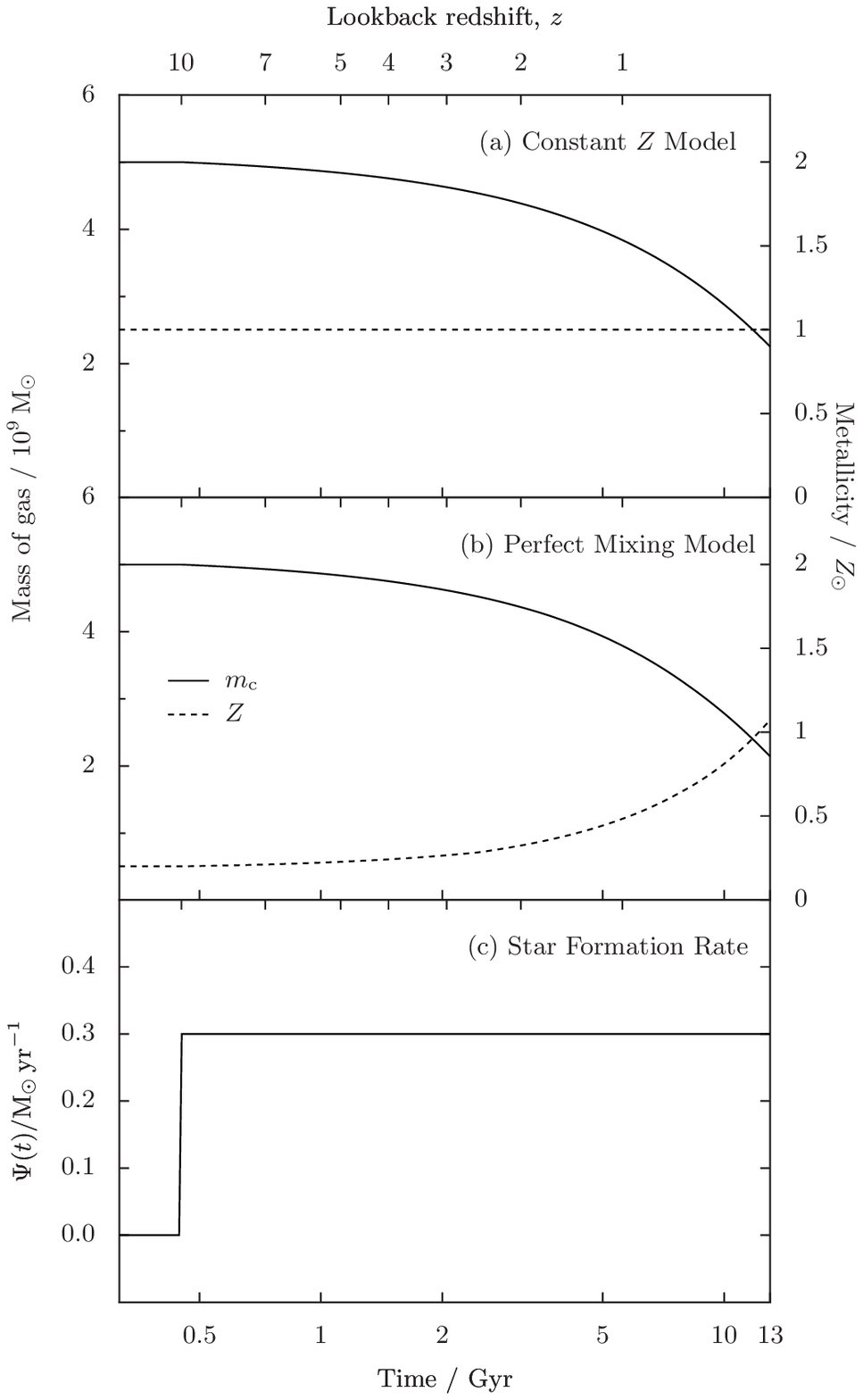}
  \caption{A model of the ISM of a galaxy of total mass $5\times 10^{9}\,\MSolar$,
  initially of metallicity 0.05 solar, but containing no stars. It begins to form stars
  at a rate of $0.3\,\Mspy$ at $z=10$, and maintains this rate of
  star formation until the current epoch. {\modified $m_\mathrm{c}$} is the total mass of gas and dust in the galaxy's {\modified ISM. Panels}~(a) and (b) represent
  differing models of the mixing of material in the ISM, as described in the
  text.}
  \label{fig:cont_closed_box}
\end{figure}

As a simple starting point, we model two dwarf-galaxy-sized clouds.  Firstly,
we model a system composed initially of $5\times 10^{9}\,\MSolar$ of gas and
dust {\modified of metallicity $0.05$}, but with no stars. This cloud begins to
undergo star formation at a rate of $0.3\,\Mspy$ at a redshift\footnote{See
Section~\ref{sec:intro} for a definition of our adopted cosmology.} of $z=10$,
and maintains this rate of star formation until the current epoch. We show the
evolution of the mass and metallicity of this system in
Figure~\ref{fig:cont_closed_box}.

\begin{figure}
  \includegraphics[width=\mnraswidth]{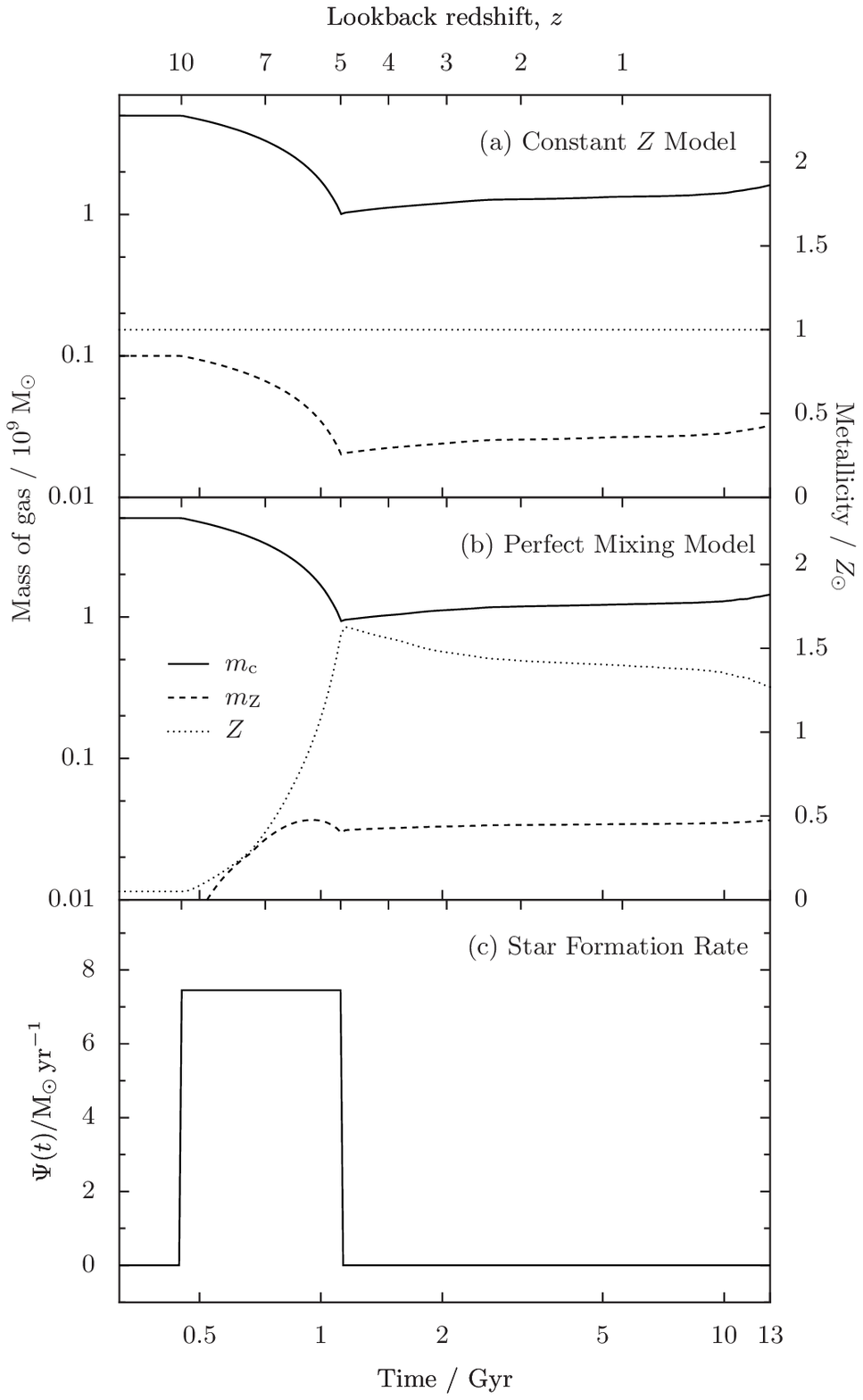}
  \caption{A model of the ISM of an early-type galaxy of total mass $5\times 10^{9}\,\MSolar$,
  which converts 90 per cent of this material into stars in the redshift range 10--5.
  {\modified $m_\mathrm{c}$} is the total mass of gas and dust in the galaxy's ISM and
  $m_\mathrm{z}$ the mass of metals. Panels~(a) and (b) represent
  differing models of the mixing of material in the ISM, as described in the
  text.}
  \label{fig:z10_5}
\end{figure}

Secondly, we form a model of an early-type galaxy.  We assume a system of total
mass $5\times 10^{9}\,\MSolar$, composed initially of zero-metallicity gas. We
assume it to undergo star formation at a constant rate in the redshift range
10--5, such that 90 per cent of this material is converted into stars over this
period. The evolution of the mass and metallicity of this system is shown in
Figure~\ref{fig:z10_5}. We note that our perfect mixing model predicts such
galaxies to have decreasing metallicities with time (Figure~\ref{fig:z10_5}b).
After $z\approx 5$, mass loss from low-mass stars dominate the return of
material to the ISM. This material has lower metallicity than supernova ejecta.

\section{Results: Variations in dust emission with physical conditions}
\label{sec:predictions}

In this section, we present model results using the diffuse geometry described
in Section~\ref{sec:rad_trans_dif}. The simple optically-thin approximation
allows us to investigate how the emissivity of dust depends upon the physical
properties of the grain population, and the radiation field which heats the
dust.  Throughout this section, we take for our interstellar heating radiation
field a \sbnn\ model of a continuous SFH of age $1\,\unit{Gyr}$, and assume the
ISM to be contained within a homogeneous spherical volume of radius $r_0$.
This leaves three free parameters: $n_\mathrm{H}$, $\chi$ and $r_0$.  We
choose, however, to re-parameterise the problem in terms of more physically
interesting quantities -- the column density of hydrogen nuclei along a
diameter of the ISM, $N_\mathrm{c}$, the total star-formation rate of the
heating radiation field, $\Psi$, and $\chi$ -- by means of the relations:
{\modified
\begin{equation}
r_0 = \sqrt \frac{\Psi L_{\nu,0}}{\chi \chi_0 c \pi} ,
\label{eq:diff_r0}
\end{equation}
}
and
\begin{equation}
n_\mathrm{H} = N_\mathrm{c} / 2r_0 ,
\end{equation}
where $L_{\nu,0}$ is the specific luminosity of a \sbnn\ model of unit SFR.
{\modified The former relation may be derived by setting the second term on the
right-hand side of Equation~(\ref{eq:diff_lum}) to equal $\Psi L_{\nu,0}$.}

We truncate our adopted ISRF at an upper energy bound $E_\mathrm{max}$, which
we set to the ionisation energy of hydrogen, $13.6\,\unit{eV}$;
astrophysically, we would expect inter-stellar neutral hydrogen to severely
attenuate the radiation field at shorter wavelengths. To avoid altering the
bolometric luminosity of the ISRF as a result of this truncation, we assume
that the truncated radiation is reprocessed to longer wavelengths with a
spectrum matching that of the nebula continuum component of the \sbnn\ model.

\begin{figure}
\includegraphics[width=\mnraswidth]{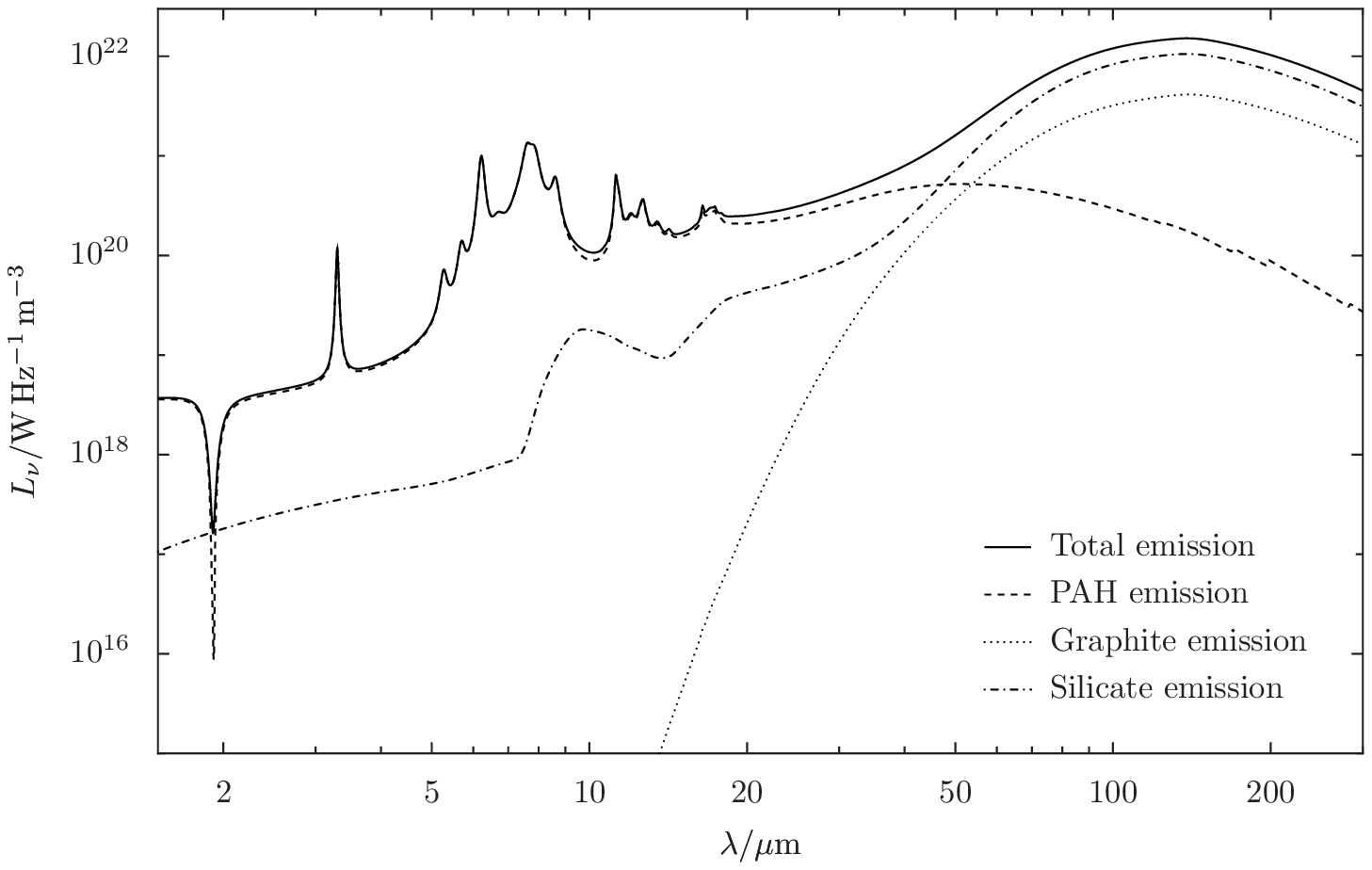}
\caption{Model spectrum of the emission from dust in our diffuse geometry,
neglecting stellar emission. We assume an ISRF appropriate for $\chi=1$,
$N_\mathrm{c}=10^{24}\,\unit{H}\,\unit{m}^{-2}$, $\Psi=1\,\Mspy$, for a system of age
$1\,\unit{Gyr}$. We decompose the total emission into the contributions from
grains composed of graphite, PAHs and silicates.}
\label{fig:gal2a}
\end{figure}

In Figure~\ref{fig:gal2a} we illustrate the contributions made by grains
composed of silicates, graphite, and PAH molecules to the total dust emission.
The division between the contributions made by graphitic and PAH grains is
somewhat blurred because of their being incorporated into a single carbonaceous
grain population with smoothly-varying optical properties. For the purposes of
the decomposition shown in Figure~\ref{fig:gal2a}, we take `graphitic' grains
to be those with radii $>50\,\unit{\AA}$, and `PAH molecules' to be those
grains with radii $<50\,\unit{\AA}$. We see that the emission shortwards of
$20\,\micron$ is dominated by grains in the carbonaceous log-normal peaks,
whilst the silicate grain population contributes around $70$ per cent of the
FIR thermal emission.  The graphitic grain population is seen to play a
relatively minor role in shaping the SED, in agreement with the finding of
\cite{2001ApJ...554..778L} that the upper limit on the sizes of the
carbonaceous grains in their model was poorly constrained.

\begin{figure}
\includegraphics[width=\mnraswidth]{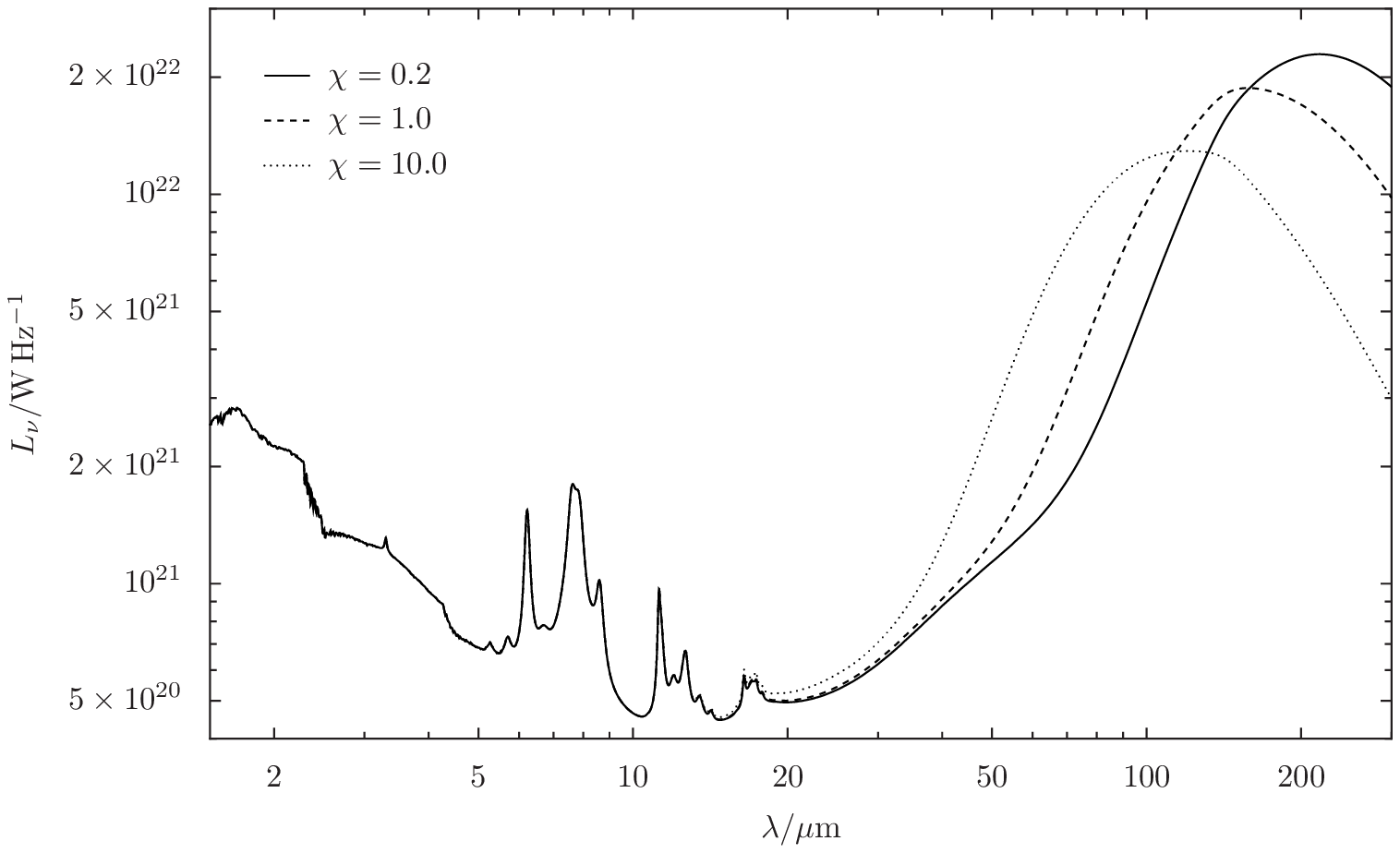}
\caption{Spectra modelled using our diffuse geometry with
$N_\mathrm{c}=10^{24}\,\unit{H}\,\unit{m}^{-2}$; we plot the sum of stellar and
dust emission. We illustrate the effect of varying the intensity $\chi$ of
the ISRF with respect to that of the solar neighbourhood.}
\label{fig:gal2b}
\end{figure}

In Figure~\ref{fig:gal2b}, we show the effect of varying the intensity $\chi$
of the ISRF with respect to that of the Solar Neighbourhood, whilst holding
$N_\mathrm{c}$ and the star-formation rate $\Psi$ of the galaxy constant.  This
means effectively altering the volume occupied by the dust and hence the energy
density in the ISRF at fixed total luminosity. As $\chi$ decreases, the peak of
the FIR thermal emission is seen to move to longer wavelengths. The temperature
of the grains contributing to this emission can be estimated from peak of the
grey body emission, showing that the temperature of the large grains falls from
about {\modified $24\,\unit{K}$ to $13\,\unit{K}$} as $\chi$ varies from $10$
to $0.2$.

In the MIR, the luminosities of the PAH features are essentially independent of
$\chi$.  This behaviour is explained by considering the transient nature of the
heating of the small grains.  For the range of conditions considered, these
small grains cool efficiently between photon absorption events.  Thus, the MIR
spectra of galaxies are independent of whether a small number of grains are
each undergoing frequent excitations, or a larger number of grains are each
undergoing less frequent excitations.  If we consider a single photon of
stellar origin, the probability of its being absorbed depends only upon the
integrated column density of dust it traverses in its passage through the ISM.
Thus, the emission of transiently-heated grains depends only upon
$N_\mathrm{c}$ and the form of the spectral energy distribution and hence
$\Psi$, independent of $\chi$.

\begin{figure}
\includegraphics[width=\mnraswidth]{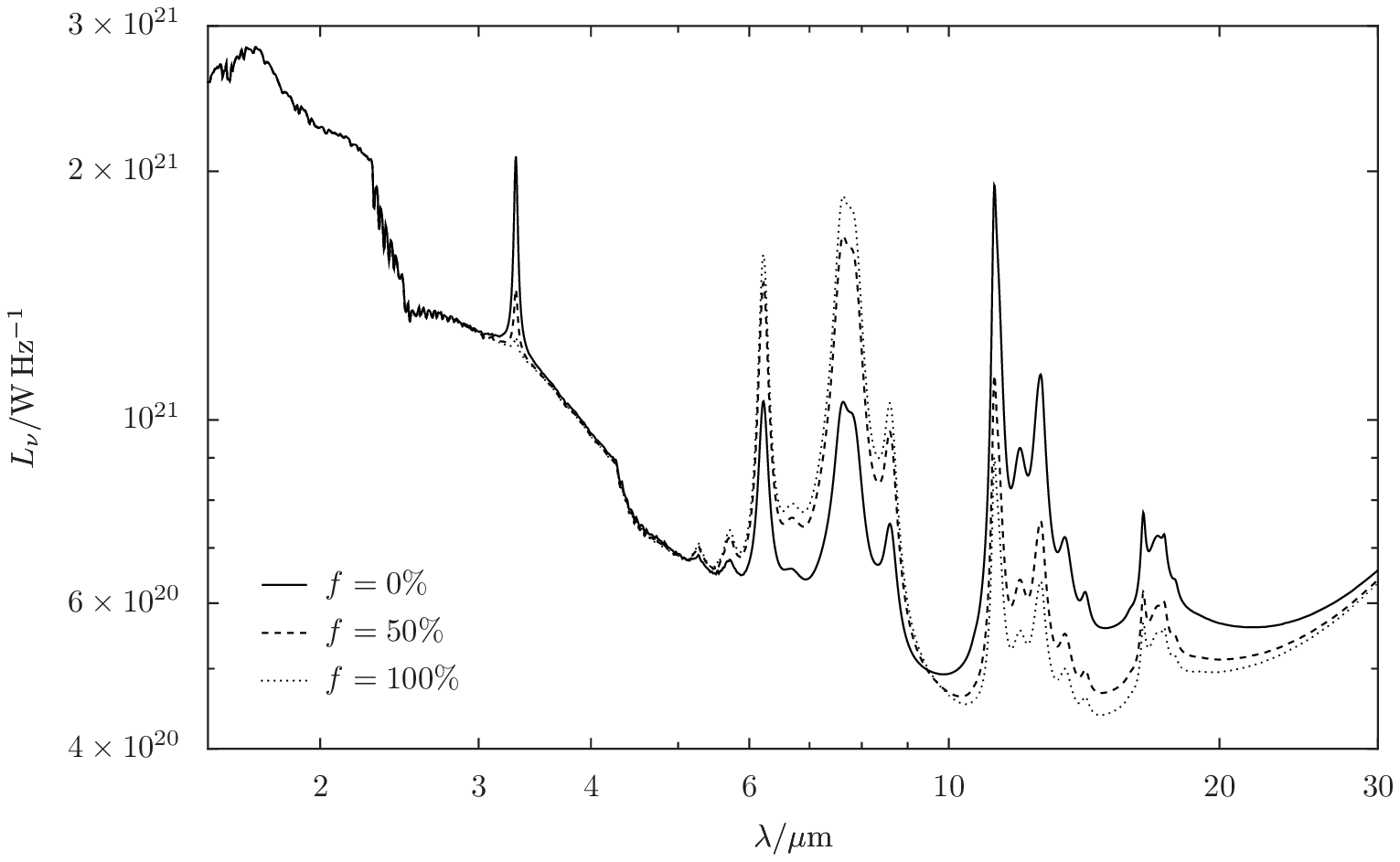}
\caption{Spectra modelled using our diffuse geometry with $N_\mathrm{c}=10^{24}\,\unit{H}\,\unit{m}^{-2}$, including both stellar and dust emission. The effect of varying the
ionisation fraction $f$ from its
default of $0.8$ is illustrated.}
\label{fig:gal2b2}
\end{figure}

In Figure~\ref{fig:gal2b2}, we show the effect of changing the
ionisation fraction $f$ of the PAH grains, as defined in
Section~\ref{sec:dust_int}, from its default of $80$ per cent. For a discussion
of observational evidence that PAH ionisation states vary between differing
astrophysical environments, see, e.g., \cite{2001ApJ...551..807D}. We see that
the MIR features in the $5 < \lambda/\micron < 9$ region are enhanced in
ionised PAHs; those in the $10 < \lambda/\micron < 20$ region are enhanced
when PAHs are neutral.

The effect of PAH ionisation can be understood in terms of the adopted
absorption cross sections shown in Figure~\ref{fig:xsec_vs_CSi}.  The
enhancement of the luminosity of the $5 < \lambda/\micron < 9$ PAH features
when PAHs are ionised can be explained simply from our adoption of a
$C^\mathrm{PAH}_{\nu,\mathrm{abs}}(a)$ which is larger at these wavelengths for
ionised PAHs; the enhancement of the $3.3\,\micron$ feature when PAHs are
neutral may similarly be explained. The enhancement of the features in the $10
< \lambda/\micron < 20$ region when PAHs are neutral, however, cannot be so
explained; $C^\mathrm{PAH}_{\nu,\mathrm{abs}}(a)$ is essentially independent of
PAH ionisation state at these wavelengths. The explanation instead lies in the
energetics of these grains.  The UV--visible absorption cross sections of PAH
molecules are essentially independent of their ionisation states; neutral and
ionised grains re-process UV--visible radiation into the MIR at essentially
identical rates. If neutral PAHs are less luminous than ionised PAHs in the $5
< \lambda/\micron < 9$ region, they must be more luminous at other wavelengths
to have the same bolometric luminosities.

\begin{figure*}
\includegraphics[width=15.5cm]{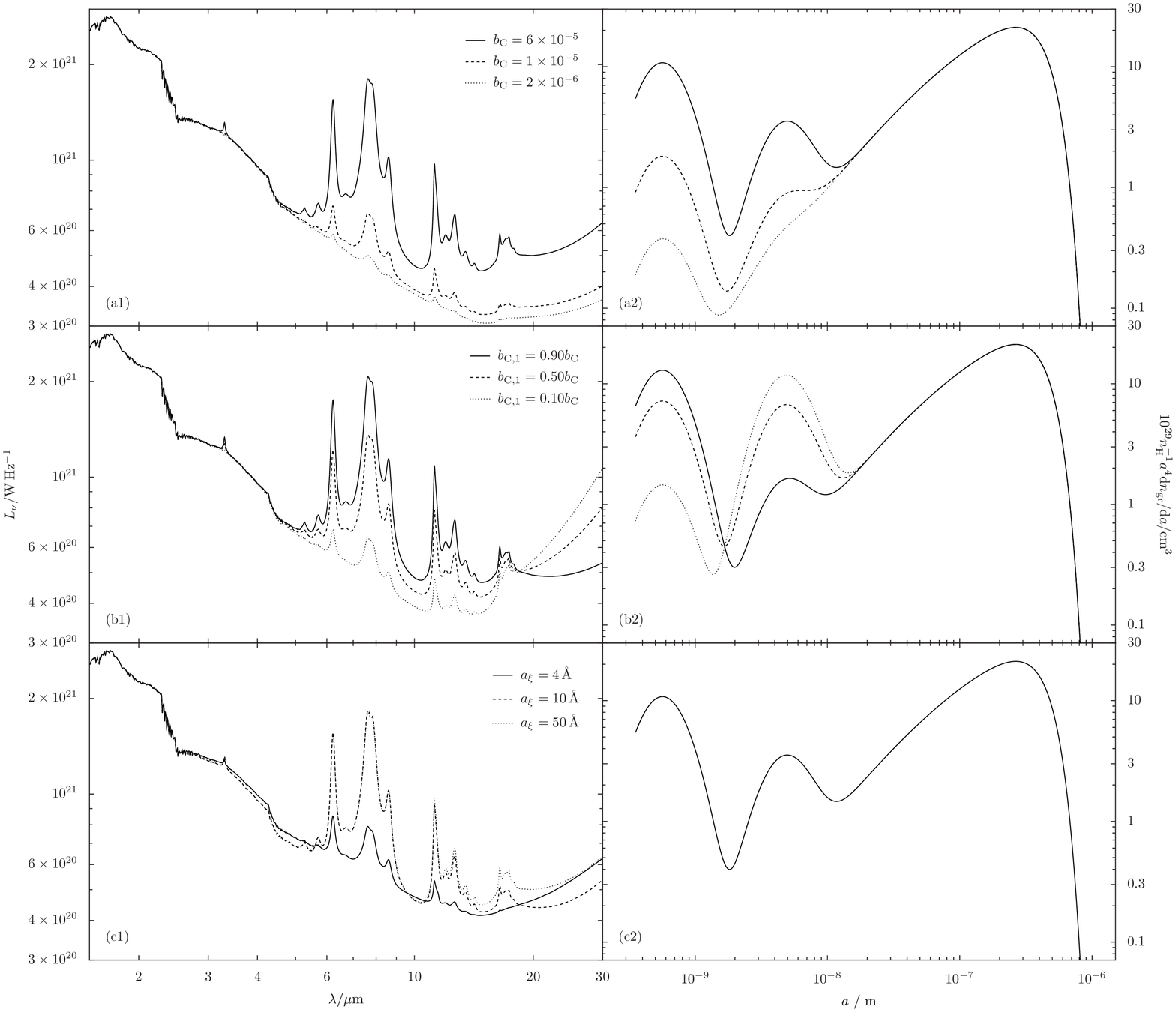}
\caption{In each row, the left plot shows spectra modelled using our diffuse
geometry with $N_\mathrm{c}=10^{24}\,\unit{H}\,\unit{m}^{-2}$, $\chi=1$ and $\Psi=1\,\Mspy$. These include both
stellar and dust emission, and illustrate the effect of varying the the size
distribution of carbonaceous grains upon the overall dust emission. The right
plot shows the size distribution of carbonaceous grains used in the plot to its left. Panel~(a1)
illustrates the effect of varying the strengths of the log-normal peaks in the
size distribution.  Panel~(b1) illustrates the effect of
varying the relative strengths of the two log-normal peaks. Panel~(c1) shows
the effect of varying the grain radius $a_\epsilon$ at which the carbonaceous
grain population makes its smooth transition from having the optical properties
of PAH molecules to those of macroscopic graphite particles.}
\label{fig:gal2c}
\end{figure*}

We now examine the variation in the MIR feature strengths as a function of the
physical properties of the population of PAH molecules.  The size distribution
of PAH grains may depend critically upon environment; in \HII\ regions, for
example, strong fluxes of ionising photons may destroy many of the smallest
grains. Such speculation is supported by the observation \citep[see,
e.g.,][]{2001ApJ...548..296W} that a diverse set of grain size distributions
are required to fit Milky Way sight lines.

In Figure~\ref{fig:gal2c}(a1) we decrease the number $b_\textrm{C}$ of
interstellar carbon nuclei per hydrogen nucleus locked up in PAH molecules from
its default value of $b_\textrm{C}=6\times 10^{-5}$. We meanwhile hold the
power-law component of the grain size distribution unchanged.  We see that the
result of removing some of the smallest grains from the model in this way is
that the MIR luminosity decreases -- eventually to leave only the Jeans tail of
the stellar emission in its place.  This result is as might be expected as
carbonaceous grains with $a<15\,\unit{\AA}$ contribute in excess of 90 per cent
of emission at $\lambda<10\,\micron$.

In Figure~\ref{fig:gal2c}(b1), the effect of altering the relative proportions
of the numbers of carbon nuclei composing PAH molecules in the two peaks is
illustrated; as defined in Section~\ref{sec:dust_pop}, $b_\mathrm{C,1}$ is the
proportion in the $3.5\,\unit{\AA}$ peak and $b_\mathrm{C,2}=1-b_\mathrm{C,1}$
that in the $30\,\unit{\AA}$ peak. Our default model has $b_\mathrm{C,1}=0.75$.
We see that the effect this change on dust emission is to enhance emission at
$\lambda\lesssim 17\,\micron$ when a greater fraction of the carbon nuclei are
placed in the $3.5\,\unit{\AA}$ peak, while reducing emission at
$\lambda\gtrsim 17\,\micron$.  A pivoting motion is seen around
$\lambda\approx17\,\micron$, where the emissivities of grains in the two
log-normal peaks, normalised per unit carbon atom, approximately equal one another.

In Figure~\ref{fig:gal2c}(c1), we consider the effect of varying $a_\xi$ -- the
grain radius, as defined in Equation~(\ref{eq:axi_transition}), at which the
optical properties of the carbonaceous grain population shift from those of
graphite to those of PAH molecules.  We see that the features at
$\lambda<11\,\micron$ are little altered when $a_\xi$ is changed from
$50\,\unit{\AA}$ to $10\,\unit{\AA}$, but that further reducing $a_\xi$ to
$4\,\unit{\AA}$ does suppress those PAH features with respect to the MIR
continuum.  As shown in Figure~\ref{fig:gal2c}(a1) above, emission at
$\lambda\lesssim 17\,\micron$ is dominated by dust in the $3.5\,\unit{\AA}$
log-normal distribution, and so is only affected once $a_\xi$ is reduced to a
comparable size. We conclude that our model predictions are relatively
insensitive to $a_\xi$, a result in agreement with \cite{2001ApJ...554..778L},
who found it to be relatively poorly constrained by observation.

\begin{figure}
\includegraphics[width=\mnraswidth]{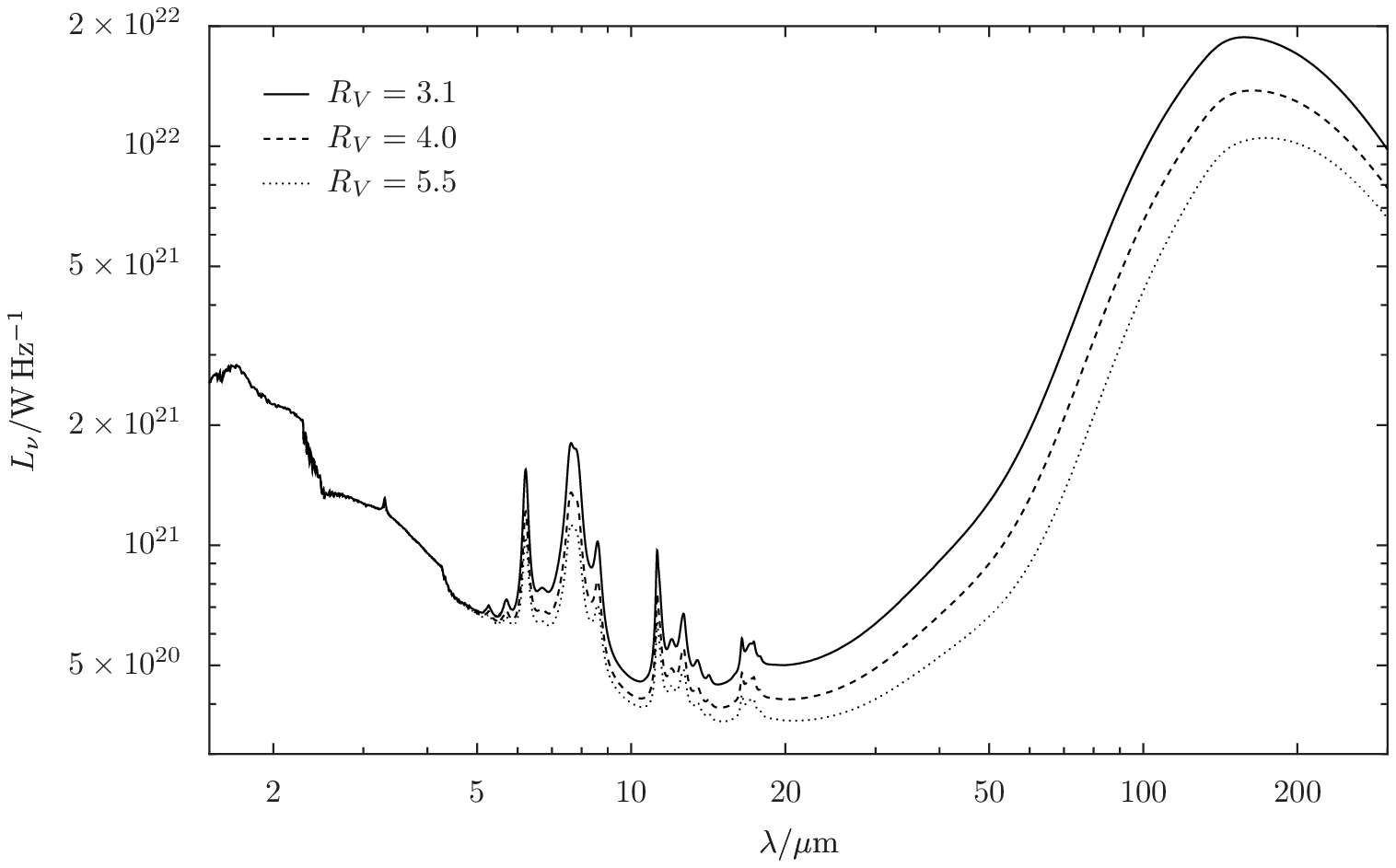}
\caption{{\modified Spectra modelled using our diffuse geometry with
$N_\mathrm{c}=10^{24}\,\unit{H}\,\unit{m}^{-2}$; we plot the sum of stellar and
dust emission. We use the grain size distributions preferred by
\citet{2001ApJ...548..296W} for Milky Way sight lines with $R_V=3.1$, $R_V=4.0$
and $R_V=5.5$. The values of $b_\mathrm{C}$ in these models are,
respectively, $6\times10^{-5}$, $4\times10^{-5}$ and $3\times10^{-5}$.
}}
\label{fig:rv_change}
\end{figure}

{\modified
Thus far we have used as our default grain size distribution that preferred by
\citet{2001ApJ...548..296W} for Milky Way sight lines with $R_V=3.1$. These are
typically sight lines which pass through infrared cirrus, away from dense
molecular clouds. However, those authors also fit grain size distributions to
the dust along sight lines which pass through more dense parts of the Milky
Way, characterised by larger values of $R_V$. For each value of $R_V$, they
present a range of fits for different assumed values of $b_\mathrm{C}$. In
Figure~\ref{fig:rv_change}, we show the spectra produced by our model when we
use the grain size distributions given by those authors for $R_V=4.0$ and
$5.5$, in each case using the value of $b_\mathrm{C}$ which they find to best
fit the sight lines studied. For these enhanced values of $R_V$, we find that
the infrared emission is reduced in line with what would be expected from the
reduced number of dust grains per unit column density in these size
distributions.
}

\section{Results: Circumnuclear shells}
\label{sec:predthick}

In this section, we present spectra modelled using our circumnuclear geometry.
As in the previous section, our default heating radiation field is a stellar
population with a continuous SFH, star-formation rate $\Psi = 1\,\Mspy$ and age
$1\,\unit{Gyr}$. We again truncate this at an upper energy bound of
$E_\mathrm{max}=13.6\,\unit{eV}$ and scale up the nebula continuum component of
the heating radiation field accordingly.

We find that the predictions of our model depend only weakly upon the
geometrical thickness, $r_1-r_0$, of the dust shell. For example, the emission
from a dust shell of inner radius $r_0 = 1\,\unit{kpc}$ and outer radius $r_1 =
3\,\unit{kpc}$ is little different from that of an {\modified infinitesimally} thin shell
at a radius of $2\,\unit{kpc}$.  Within the accuracy of our numerical
integration, these two models are indistinguishable at $\lambda \lesssim
14\,\micron$. At longer wavelengths, they do differ, the thicker shell having
an enhanced luminosity at 15--$70\,\micron$, enhanced by 23, 72 and 29 per cent
at 20, 40 and $60\,\micron$ respectively. This can be understood in terms of
two effects: an enhanced probability of multiple-photon heating of small grains
close to the inner edge of the shell, and radial variations in the temperatures
of the large dust grains. For the remainder of this section, we adopt a
geometry with $r_0 = 1\,\unit{kpc}$ and $r_1 = 3\,\unit{kpc}$.

\begin{figure}
\includegraphics[width=\mnraswidth]{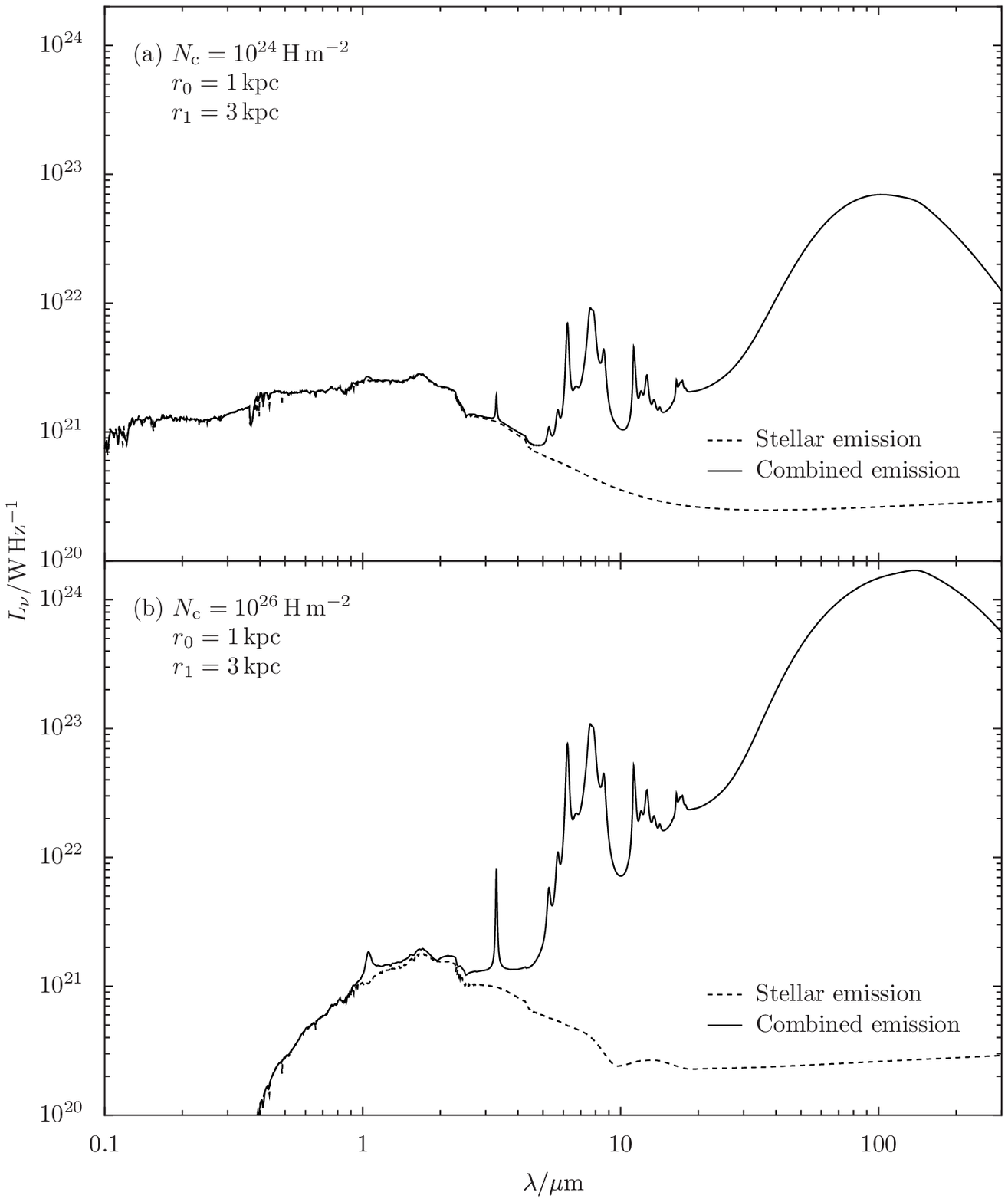}
\caption{Spectra of dusty star-forming regions modelled using our circumnuclear
geometry. For the heating radiation field, a stellar population with continuous
SFH, star-formation rate $\Psi = 1\,\Mspy$ and age
$1\,\unit{Gyr}$ is used. Panels~(a) and (b) show spectra of dust
shells of column densities $N_\mathrm{c}=10^{24}$ and $10^{26}\,\unit{m}^{-2}$
respectively.}
\label{fig:gal2d2}
\end{figure}

Figure~\ref{fig:gal2d2} shows model spectra for dust shells with column
densities $N_\mathrm{c}=10^{24}$ and $10^{26}\,\unit{H}\,\unit{m}^{-2}$; we
assume solar metallicity for both the stellar populations and the dust shells.
For each, we show the contribution made by attenuated stellar radiation -- i.e.
$I_\nu^{(1)}$ in Equation~(\ref{eq:rr_decomp}) -- given by:
\begin{equation}
L_\mathrm{stellar} = L_\nu e^{-C_{\nu,\mathrm{abs}} N_\mathrm{c}} ,
\end{equation}
where $L_\nu$ is the luminosity of the central heating source.

When $N_\mathrm{c} = 10^{24}\,\unit{H}\,\unit{m}^{-2}$, stellar emission
contributes 95 per cent of emission at $4\,\micron$. Convolution of this
spectrum with the passbands of {\it Spitzer}'s InfraRed Array Camera (IRAC)
reveals that stellar emission contributes 92, 91, 34 and 11 per cent in the
channels at $3.5$, $4.5$ $5.8$ and $8.0\,\micron$ respectively.

At $N_\mathrm{c} = 10^{26}\,\unit{H}\,\unit{m}^{-2}$, these contributions
reduce to 60 per cent of emission at $4\,\micron$, and 48, 45, $3.9$ and $0.88$
per cent in the IRAC channels at $3.5$, $4.5$, $5.8$ and $8.0\,\micron$
respectively.  An absorption feature in the attenuated stellar emission
component is seen around $10\,\micron$, indicating that the silicate
$9.7\,\micron$ feature has an optical depth of $0.4$ in this model.

In these model spectra, two PAH features appear well-isolated from their
neighbours: that at $3.3\,\micron$ and the new $1.05\,\micron$ feature added by
DL07.  In the $N_\mathrm{c} = 10^{26}\,\unit{H}\,\unit{m}^{-2}$ model, these
have equivalent widths $0.3$ and $0.04\,\micron$ respectively. In the
$N_\mathrm{c} = 10^{24}\,\unit{H}\,\unit{m}^{-2}$ model, the $3.3\,\micron$
feature has equivalent width $0.07\,\micron$; the $1.05\,\micron$ feature is
not apparent.

\begin{figure}
\includegraphics[width=\mnraswidth]{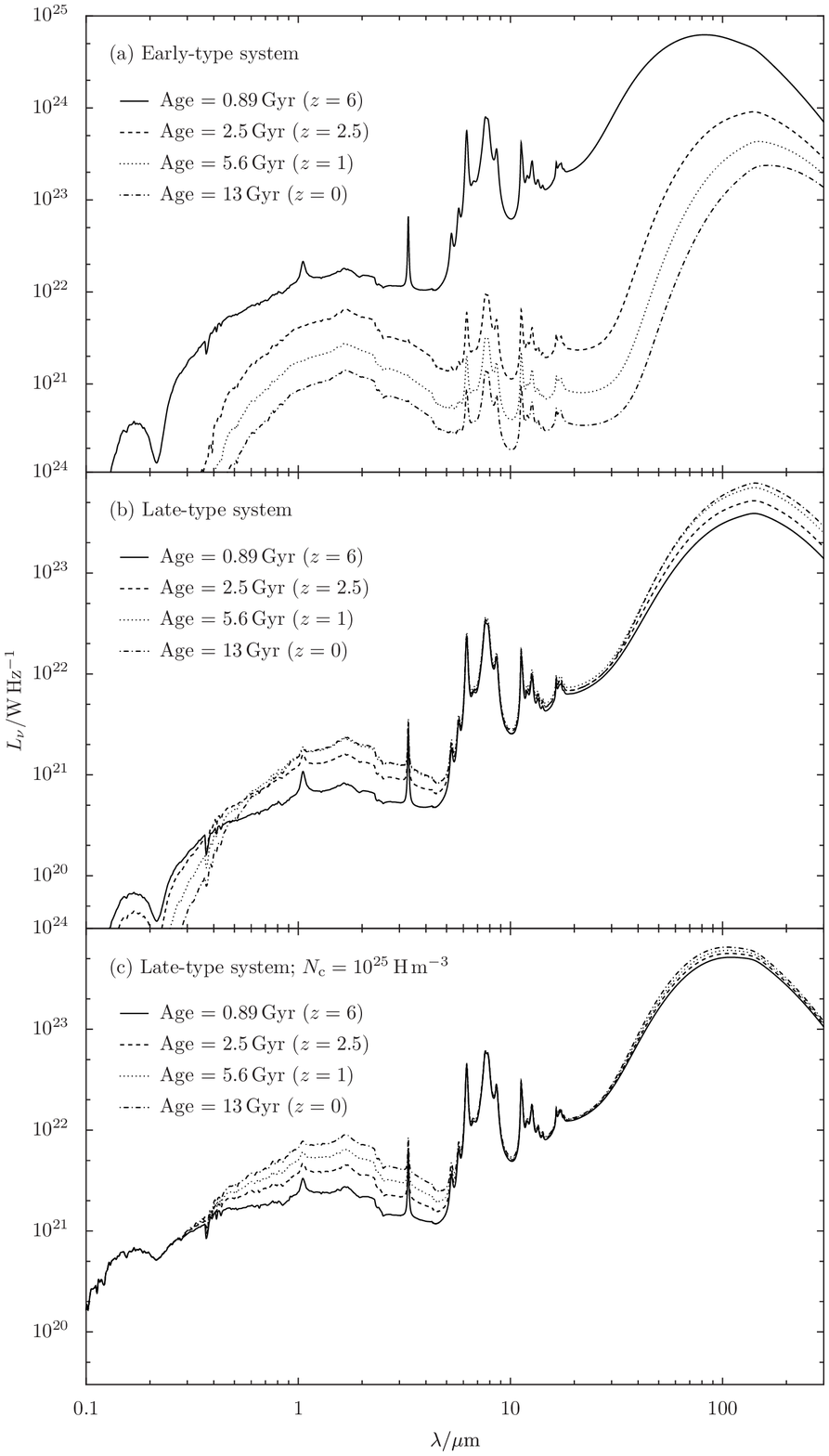}
\caption{The time-evolution of the spectra of dust-enshrouded star-forming
regions modelled using our circumnuclear geometry. In all cases, $r_0 =
1\,\unit{kpc}$ and $r_1 = 3\,\unit{kpc}$. Panel~(a) shows the evolution of the SED
of the early-type galaxy modelled in Figure~\ref{fig:z10_5}; the star-formation
history and dust column density assumed are as shown in that figure for the
perfect mixing model. Panel~(b) shows the evolution of the system with ongoing
star formation modelled in Figure~\ref{fig:cont_closed_box}. Panel~(c) shows a
system with the same SFH as Panel~(b), but assuming a constant column density
$N_\mathrm{c}=10^{25}\,\unit{H}\,\unit{m}^{-2}$.}
\label{fig:gal2e2}
\end{figure}

{\modified
In these plots we have, as previously, assumed the grain size distribution
which \citet{2001ApJ...548..296W} fit to Milky Way sight lines with $R_V=3.1$,
assuming $b_\mathrm{C}=6\times10^{-5}$. From Figure~\ref{fig:rv_change}, we can
see that the effect of using a size distribution appropriate for a larger value
of $R_V$ would be a small reduction in dust emission at all wavelengths. 
}

In Figure~\ref{fig:gal2e2} we trace the evolution of our model spectra as a
function of age for a range of SFHs. Figure~\ref{fig:gal2e2}(a) traces the
evolution of the early-type galaxy modelled in Figure~\ref{fig:z10_5};
Figure~\ref{fig:gal2e2}(b) traces that of the system with ongoing star
formation modelled in Figure~\ref{fig:cont_closed_box}.

Figure~\ref{fig:gal2e2}(c) shows a system with the same SFH as that modelled in
Figure~\ref{fig:gal2e2}(b), but assuming a constant dust shell column density
$N_\mathrm{c}=10^{25}\,\unit{H}\,\unit{m}^{-2}$.  The time-invariance of the emission at
$\lambda\lesssim 300\,\unit{nm}$ can be understood in terms of an equilibrium
being reached between star formation and stellar death among the massive stars
which produce it.  This contrasts with the behaviour seen in
Figure~\ref{fig:gal2e2}(b), where the system becomes more metal-rich and dusty
as it evolves, increasing extinction at ultraviolet wavelengths.

\section{The dependence of MIR luminosity upon SFR}
\label{sec:validation}

To probe the dependence of the MIR luminosities of our models upon SFR, we
constructed a set of circumnuclear models, each having a continuous SFH of age
$1\,\unit{Gyr}$ but different star formation rates, ranging from
0.02 to $100\,\Mspy$. We adopt otherwise constant model parameters: $r_0 =
1\,\unit{kpc}$, $r_1 = 3\,\unit{kpc}$ and $N_\mathrm{c} =
10^{25}\,\unit{H}\,\unit{m}^{-2}$. We constructed two sets of models, using the
LD01 and the DL07 PAH cross sections, to compare their predictions.

We compare these to the sources in the {\it Spitzer} extragalactic First Look
Survey (FLS) field. We correlated this source catalogue with the Fourth Data
Release of the Sloan Digital Sky Survey
\citep[SDSS;][]{2000AJ....120.1579Y,2006ApJS..162...38A} using a $2''$ matching
radius to obtain redshifts for these sources, and used emission line
measurements from the MPA/JHU catalogue\footnote{Available from:
\url{http://www.mpa-garching.mpg.de/SDSS/}}.  To remove AGN and
spurious/marginal detections from the catalogue, and to correct for aperture
effects and extinction, we followed \cite{2004MNRAS.355..874N}: AGN were
identified using the spectral classification scheme of
\cite{1987ApJS...63..295V}; aperture and extinction corrections were applied to
the emission line data using the prescription of \cite{2003ApJ...599..971H}.
We estimated the SFRs of the sources by comparing their \Halpha\ luminosities
with that predicted by \sbnn\ for a $1\,\unit{Gyr}$-old continuous SFHs.
$K$-corrections for the {\it Spitzer} $8\,\micron$ and $24\,\micron$ fluxes
were derived using our models described above, selecting for each object that
model which most closely matched the \Halpha-derived SFR.

\begin{figure*}
  \includegraphics[clip,width=15.5cm]{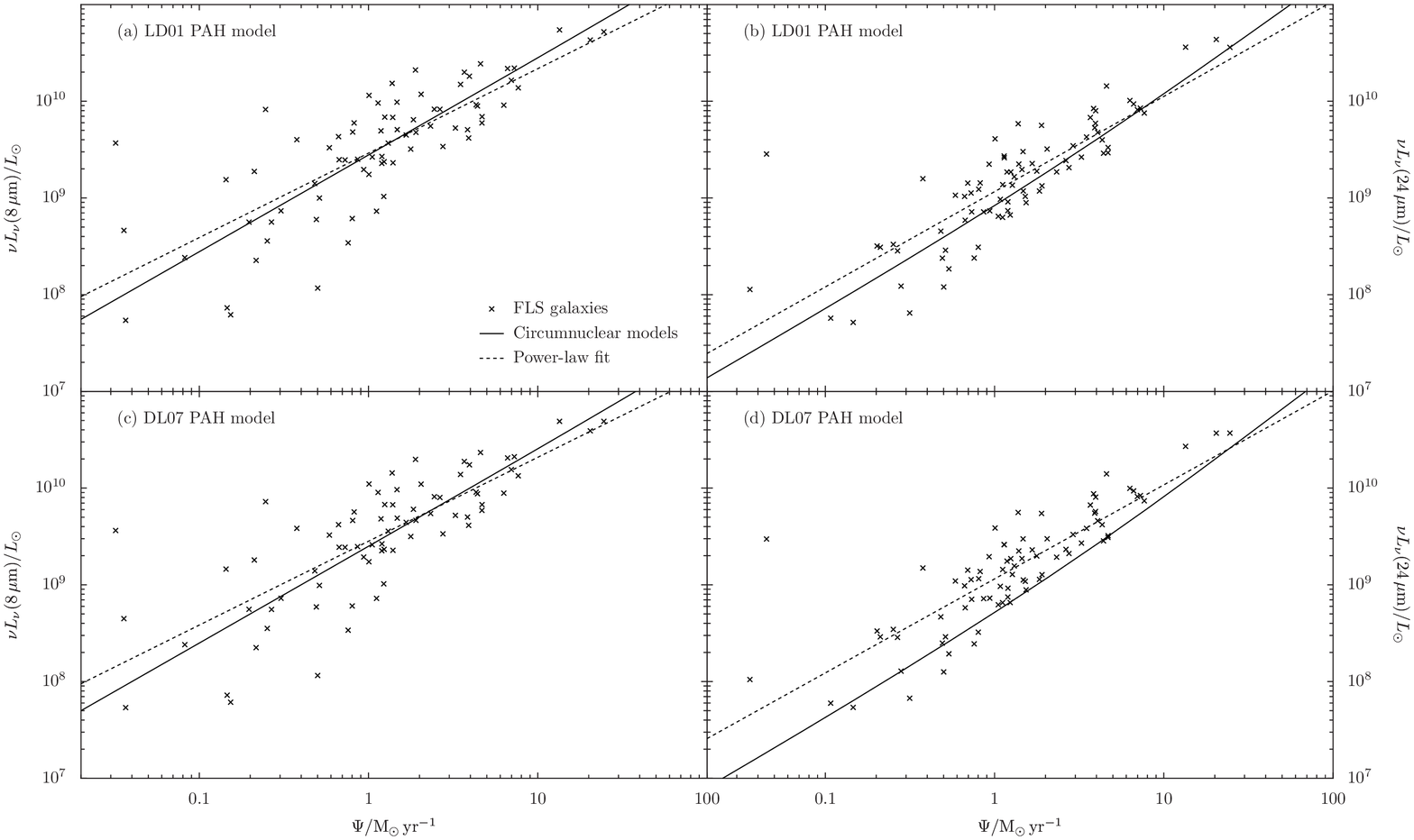}
  \caption{Solid lines: The {\it Spitzer\/}-convolved 8-$\micron$ (Left panels)
     and 24-$\micron$ (Right panels) luminosities of circumnuclear models with
	  continuous SFHs, age $1\,\unit{Gyr}$, $r_0 = 1\,\unit{kpc}$, $r_1 =
	  3\,\unit{kpc}$ and $N_\mathrm{c} = 10^{25}\,\unit{H}\,\unit{m}^{-2}$.
	  Dashed lines: Power-law fits to the {\it Spitzer} extragalactic FLS
     source population (points).  In the upper panels we use the LD01 PAH cross
     sections; in the lower panels we use the DL07 cross sections.}
  \label{fig:flscorrs}
\end{figure*}

We convolved our model spectra with the 8-$\micron$ and 24-$\micron$ passbands
of {\it Spitzer} to produce matching data from our model and the observed
source population.  The results are shown in Figure~\ref{fig:flscorrs}; also
shown are power-law fits to the observed source population, the fit
coefficients for which are listed in Table~\ref{tab:wu_powerlaw} \citep[c.f. a
similar fit performed by][]{2005ApJ...632L..79W}. In the upper panels, we use
the LD01 PAH cross sections (see Section~\ref{sec:dust_int}) in our models; in
the lower panels we use the DL07 cross sections. The non-linearity of our model
$L_\nu(24\,\micron)-\Psi$ relations at $\Psi>10\,\Mspy$ results from the
increasing emission of grains excited by multiple photons at the highest star
formation rates.

{\modified
These results are insensitive to the chosen geometry, as the emission from our
models at both 8 and $24\,\micron$ is dominated by grains in the transiently
heated regime. The luminosities are, however, dependent upon the column density
of dust; this is likely to be one cause of the scatter of the FLS sources from
our model relations. Increasing $N_\mathrm{c}$ by an order of magnitude, to
$10^{26}\,\unit{H}\,\unit{m}^{-2}$, increases our predictions of
$L_\nu(8\,\micron)$ by a factor of around 1.75, and $L _\nu(24\,\micron)$ by a
factor of around 2. Increasing $N_\mathrm{c}$ further causes our predictions
for $L_\nu(8\,\micron)$ to decrease as self-absorption becomes significant.
Further scatter in these plots is likely to result from the uncertainly in the
conversion of \Halpha\ luminosities into star formation rates.
}

It is interesting to note that a better fit to the 24-$\micron$ luminosities of
the FLS sources is obtained using the LD01 PAH cross sections compared to the
DL07 cross sections (see Figures~\ref{fig:flscorrs}b and \ref{fig:flscorrs}d).
This difference can be explained by the enhanced PAH cross sections at
$24\,\micron$ in the LD01 model, as seen in Figures~\ref{fig:xsec_vs_CSi}(a)
and \ref{fig:xsec_vs_CSi}(b).  The fit to the $24\,\micron$ luminosities in
Figure~\ref{fig:flscorrs}(d) can be improved by increasing $N_\mathrm{c}$, but
this leads to an over-prediction of the 8-$\micron$ luminosities in
Figure~\ref{fig:flscorrs}(c).

\begin{table}
\begin{center}
\begin{tabular}{llll}
\hline
{$\lambda$} & {$a$} & {$b$} & {$N$} \\
\hline
$ 8\,\micron$ & $9.45\pm0.05$ & $0.87\pm0.08$ & 75 \\
$24\,\micron$ & $9.06\pm0.04$ & $0.97\pm0.07$ & 81 \\
\hline
\end{tabular}
\end{center}
\caption{Coefficients of the power-law fits shown
	in Figure~\ref{fig:flscorrs}, where $\log_{10} \left( \nu L_\nu (\lambda) /
   L_\odot \right) = a + b \times \log_{10} \left( \Psi / \Mspy \right)$. The
   number of sources used in each fit is $N$. Although the $K$-corrections
   derived for the FLS galaxies differ depending upon whether the LD01 or DL07 PAH cross sections are used, these
   differences are modest; the changes to the power-law coefficients above are much smaller than the quoted errors.}
\label{tab:wu_powerlaw}
\end{table}

\section{Predicted Galaxy Colours}
\label{sec:colours}

\begin{figure}
\includegraphics[width=\mnraswidth]{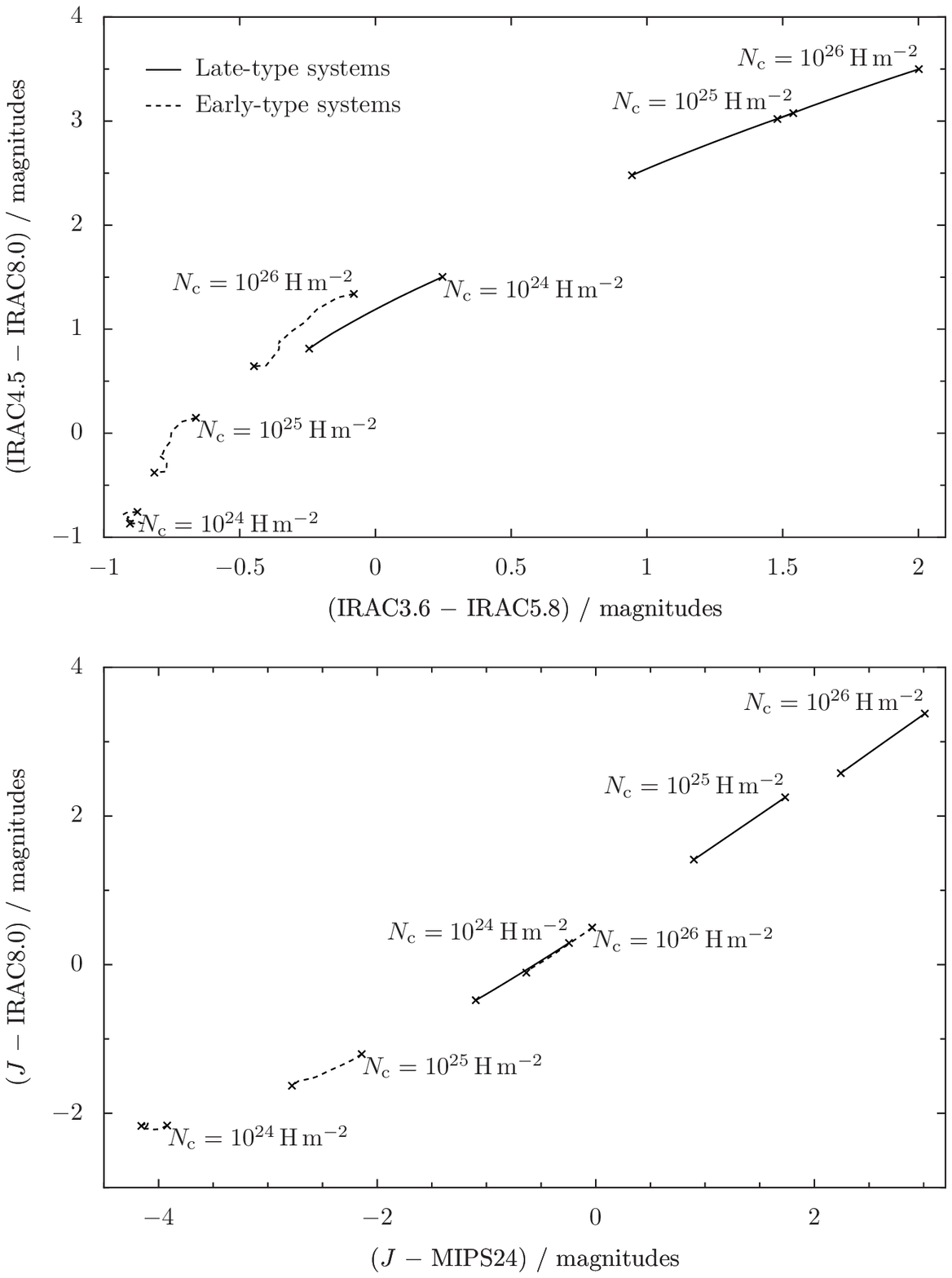}
\caption{\modified Plot of the positions of galaxies modelled by our circumnuclear geometry in colour-colour space. Each trace represents the evolution of a model galaxy from an age of $1.6\,\mathrm{Gyr}$ (top-right end) to $13.8\,\mathrm{Gyr}$ (lower-left end). Late-type galaxies, with continuous SFHs, are shown with solid traces; early-type galaxies, with instantaneous SFHs, are shown with dashed traces. Three traces are shown for each galaxy type, representing galaxies enshrouded with three different column densities of dust. All colours are shown in magnitudes, and are calculated in the galaxy rest frame.}
\label{fig:colcol1}
\end{figure}

{\modified Using the models presented in Section~\ref{sec:predthick}, we can
make predictions of the relative merits of colour diagnostics in the {\modified
discrimination} of systems with ongoing star formation from those with only old
stellar populations. In Figure~\ref{fig:colcol1}, we show the colours of late-
and early-type models, with continuous and instantaneous SFHs respectively,
enshrouded by three different column densities of dust.  In} the lower panel we
consider colours formed between the $J$-band of the UK InfraRed Telescope
(UKIRT), as used by the UKIRT Infrared Deep Sky Survey (UKIDSS), in conjunction
with {\it Spitzer} MIR photometry. In the upper panel, we consider colours
which use only IRAC photometry. In all cases, the late-type models appear
redder than systems with only old stellar populations, a result of the enhanced
UV-pumping of PAH grains. {\modified Models which are enshrouded by higher
column densities of dust appear systematically redder, as a result of their
enhanced dust emission.}

\begin{figure*}
\includegraphics[width=15.5cm]{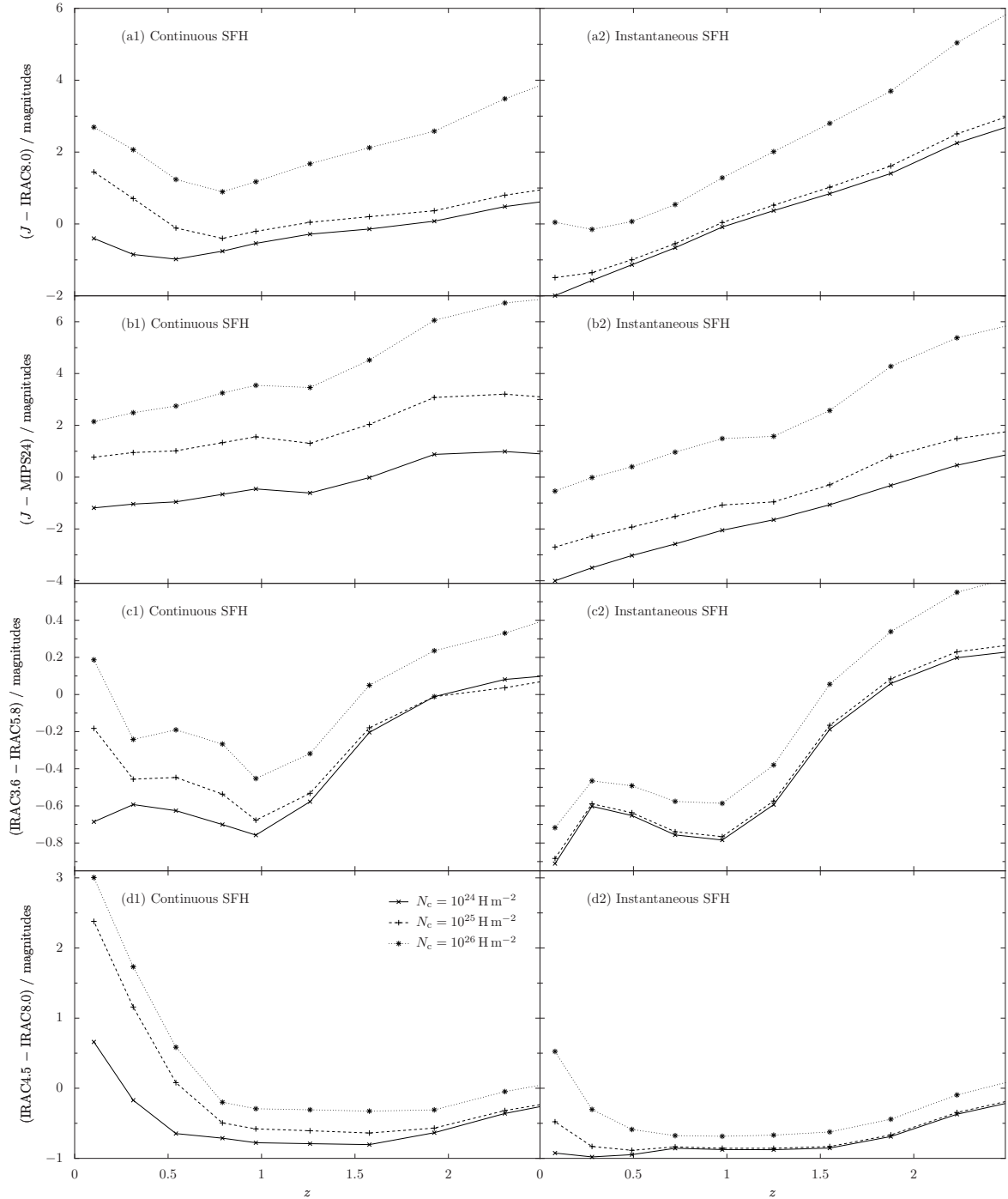}
\caption{The colours of galaxies modelled by our circumnuclear geometry with continuous (left panels) and instantaneous (right panels) SFHs; in each case, star formation starts at $z=10$. Each trace represents the redshift evolution of a system of constant $N_\mathrm{c}$. All colours are shown in magnitudes, and are calculated in the frame of the observer.}
\label{fig:colcol2}
\end{figure*}

All of these colour predictions are robust to changes in star-formation rate
$\Psi$. In our model, both stellar and PAH emission are linearly proportional
to $\Psi$ except at the highest star formation rates (see
Figure~\ref{fig:flscorrs}); only the far-infrared greybody varies in colour
with $\Psi$.  {\modified These predictions are also robust to changes in the
assumed masses of the galaxies if $\Psi$ is assumed to scale linearly with mass
whilst $N_\mathrm{c}$ remains constant.} In Figure~\ref{fig:colcol2} we
investigate the redshift and column density dependence of the colours shown in
Figure~\ref{fig:colcol1} by tracing the redshift evolution of systems of
constant $N_\mathrm{c}$. The diagnostic power of colours formed between IRAC
channels is limited beyond $z\gtrsim 0.5$; at higher redshifts these bands are
increasingly dominated by stellar emission, especially at lower dust column
densities. The power of $J-8\,\micron$ and $J-24\,\micron$ colours extends to
higher redshifts, though a degeneracy exists between highly obscured old
stellar populations and recent star formation. In all cases, the two SFHs
considered in Figure~\ref{fig:colcol2} converge at high redshift; both contain
relatively young stellar populations shortly after their formation at $z=10$.

\section{Summary}

We have developed a semi-empirical model of the emission of dust in
star-forming galaxies, capable of producing model spectra which extend from the
Lyman limit through three orders of magnitude in wavelength to the
far-infrared. A particular strength is its state-of-the-art model of the
mid-infrared PAH features: we treat the transient heating of small dust grains
using the sophisticated method of \cite{2001ApJ...551..807D} and use the grain
interaction cross sections of \cite{2007ApJ...657..810D}.

In Section~\ref{sec:predictions}, we showed how the predictions of our model
depend upon the microscopic properties of the dust grain population -- its
grain size spectrum and ionisation state, for example. In
Sections~\ref{sec:predthick} and \ref{sec:validation}, we went on to
demonstrate the spectra produced by our model depend upon the large scale
properties of galaxies: their star formation histories and dust column
densities. In Figure~\ref{fig:flscorrs}, we found that the {\it Spitzer}
extragalactic First Look Survey (FLS) source population is better matched by
models which use the PAH cross sections of \cite{2001ApJ...554..778L} than
those which use the cross sections of \cite{2007ApJ...657..810D}. 

In Section~\ref{sec:colours}, we provide estimates of the near-to-mid-infrared
colours of early- and late-type galaxies, demonstrating the power of our model
as a tool for interpreting colour-colour diagrams.

\section*{Acknowledgments}

DCF acknowledges the receipt of a PPARC studentship. This work has made use of
the distributed computation grid of the University of Cambridge
(\textsc{CamGRID}). {\modified We are grateful to the anonymous referee
for helpful comments.}

\appendix

\section{Choice of energy bins}

The energy bins used in Section~\ref{sec:transient_heat} are chosen so that
$P(E)$ is well-sampled in regions where it is changing rapidly.  For the
smallest grains, this means that the ground state, and those close to it, must
be well-sampled. To ensure that this is the case, we set the $N_1$
lowest-energy bins in our scheme to contain the $N_1$ lowest-energy vibrational
states of dust grains, one in each bin, thus treating them exactly. At the
opposite extreme, we require the occupation probability of our highest-energy
bin to be very small, such that the high-energy tail of $P(E)$ is not
truncated.

To meet these requirements, we place the next $N_2$ bins at linear intervals of
$\boldsymbol{U}_{N_1} - \boldsymbol{U}_{N_1-1}$ above $\boldsymbol{U}_{N_1}$.
We then place the remaining $N_\mathrm{bins}-N_1-N_2$ bins at logarithmic
intervals above $\boldsymbol{U}_{N_1+N_2}$ with multiplicative spacing $\alpha$
given by:

\begin{equation}
\alpha = \left\{ \begin{array}{llllll}
1.1 & & & a & < & 10\,\mathrm{\AA} \\
1.5 & 10\,\mathrm{\AA} & \leq & a & < & 50\,\mathrm{\AA} \\
1.5 + \frac{a - 50\,\mathrm{\AA}}{350\,\mathrm{\AA} - 50\,\mathrm{\AA}}(7.0 - 1.5) & 50\,\mathrm{\AA} & \leq & a & < & 350\,\mathrm{\AA} \\
7.0 & 350\,\mathrm{\AA} & \leq & a & & \\
\end{array}\right.
\end{equation}

In this paper, we use $N_1=20$, $N_2=50$ and $N_\mathrm{bins}=600$.

\section{Calculating the transition matrix $T$}

Several approximations for calculating the emission from
stochastically heated dust grains are discussed in detail by
\cite{2001ApJ...551..807D}. They all rely on dividing the possible
enthalpy content of the grain into of the order of 500 bins and
calculating the probability of a grain transitioning from an enthalpy
that belongs to one bin to an enthalpy belonging to a different bin.

\begin{figure}
\begin{centering}
\begin{pspicture}(0,0)(10,5)
  \psframe[fillstyle=solid,fillcolor=lightgray,linewidth=0pt](0,0.0)(8,1.5)
  \psline[linewidth=2pt](0,0.0)(8,0.0)\uput[0](8,0.0){$U_{i}^\text{min}$}
  \psline[linewidth=2pt](0,1.5)(8,1.5)\uput[0](8,1.5){$U_{i}^\text{max}$}
  \psframe[fillstyle=solid,fillcolor=lightgray,linewidth=0pt](0,3.5)(8,4.5)
  \psline[linewidth=2pt](0,3.5)(8,3.5)\uput[0](8,3.5){$U_{j}^\text{min}$}
  \psline[linewidth=2pt](0,4.5)(8,4.5)\uput[0](8,4.5){$U_{j}^\text{max}$}
  \psline[linewidth=1pt]{<->}(0.5,1.5)(0.5,3.5)\uput[0](0.5,2){$W_{1}$}
  \psline[linewidth=1pt]{<->}(2.0,1.5)(2.0,4.5)\uput[0](2.0,2){$W_{2}$}
  \psline[linewidth=1pt]{<->}(3.5,0)(3.5,3.5)\uput[0](3.5,2){$W_{3}$}  
  \psline[linewidth=1pt]{<->}(5.0,0)(5.0,4.5)\uput[0](5.0,2){$W_{4}$}  
  \psline[linewidth=1pt]{<->}(7,0.0)(7,1.5)\uput[0](7,0.75){$\Delta U_{i}$}  
  \psline[linewidth=1pt]{<->}(7,3.5)(7,4.5)\uput[0](7,4.00){$\Delta U_{j}$}  
\end{pspicture}
\end{centering}
\caption{Illustration of the energy boundaries used in function
  $\mathbf{G}_{ji}(E)$, where $j$ and $i$ are the
  indices of the upper and lower bins respectively and $\boldsymbol{U}_{i}^{\rm min}$
  and $\boldsymbol{U}_{i}^{\rm max}$ are the boundaries of the $i$th bin. Note that
  $W_{2}$ is the smaller and $W_{3}$ the larger of the pair of values:
  $(\boldsymbol{U}_{i}^{\rm max} - \boldsymbol{U}_{j}^{\rm max})$ and $(\boldsymbol{U}_{i}^{\rm min}-
  \boldsymbol{U}_{j}^{\rm min})$. }
\label{fig:Wexplain}
\end{figure}
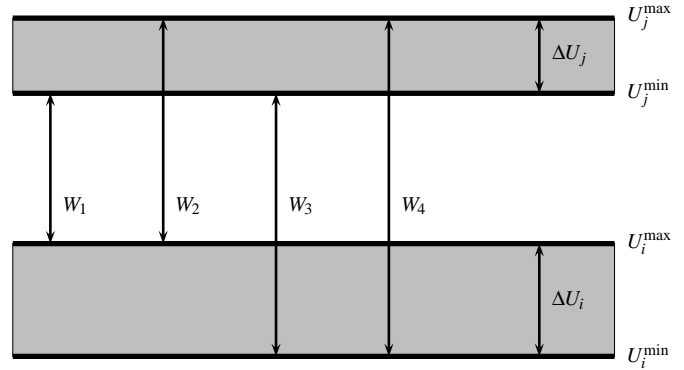

{\modified
The upward transition rates $\mathbf{T}_{ji}$ are given
by \citep[][equation~15]{2001ApJ...551..807D}:
\begin{equation}
\label{eq:upwardtrans}
  \mathbf{T}_{ji} = \frac{ c \Delta \boldsymbol{U}_{j} }{\boldsymbol{U}_{j}-\boldsymbol{U}_{i}} \int_{W_{1}}^{W_{4}}
  \mathbf{G}_{ji} (E) C_{\rm abs}(E) u_{E} (E) \,\mathrm{d}E \qquad \text{for $j<N_\mathrm{bins}$,}
\end{equation}
\begin{eqnarray}
\mathbf{T}_{ji} & = & \frac{c}{\boldsymbol{U}_{j}-\boldsymbol{U}_{i}} \left[
\int_{W_1}^{W_\mathrm{c}} \left(\frac{E-W_1}{W_\mathrm{c}-W_1}\right)
C_{\rm abs}(E) u_{E} (E) \,\mathrm{d}E \right. \nonumber \\
& & \left.+\int_{W_\mathrm{c}}^{\infty} C_{\rm abs}(E) u_{E} (E) \mathrm{d}E \right]  \qquad \text{for $j=N_\mathrm{bins}$}
\end{eqnarray}
where $C_{\rm abs}$} is the grain absorption cross section, $u_{E}$ is
the energy density of the radiation field and $\mathbf{G}_{ji}$ is a correction
factor for the finite width of bins and is given by:
\begin{equation}
  \mathbf{G}_{ji} = \left\{
    \begin{aligned}
       \frac{E-W_{1}}{\Delta \boldsymbol{U}_{j}\Delta \boldsymbol{U}_{i}} && W_{1}<E<W_{2} ,\\
       \frac{ {\rm min} \left( \Delta \boldsymbol{U}_{j} , \Delta \boldsymbol{U}_{i} \right)
       }{\Delta \boldsymbol{U}_{j}\Delta \boldsymbol{U}_{i}} && W_{2}<E<W_{3}, \\
       \frac{W_{4}-E}{\Delta \boldsymbol{U}_{j}\Delta \boldsymbol{U}_{i}} &&
       W_{3}<E<W_{4},\\
       0 &&\text{otherwise.}
  \end{aligned}
  \right.
\end{equation}
The integration limit quantities $W_{1}$, $W_{2}$, $W_{3}$ and $W_{4}$
are defined in Figure~\ref{fig:Wexplain}. {\modified $W_\mathrm{c}$ is equal to
$U_{N_\mathrm{bins}}^{\mathrm{min}} - U_i^{\mathrm{min}}$.}

{\modified
Throughout this paper, we use the continuous cooling approximation for
the downward transitions \citep[][equation~41]{2001ApJ...551..807D}:
\begin{eqnarray}
\mathbf{T}_{ji} & = & \frac{1}{\boldsymbol{U}_{i}-\boldsymbol{U}_{j}}
\frac{8\pi}{h^3 c^2} \nonumber \\
& & \times \int_{0}{E_u} \frac{E^3 C_{\rm abs}(E)}{\exp(E/k\boldsymbol{\theta}_u)-1}\,\mathrm{d}E
\qquad \text{for $i>1$; $j=i-1$}
\label{t_downward}
\end{eqnarray}
\begin{equation}
\mathbf{T}_{ji} = 0 \qquad \text{for all other $\{i,j\}$}
\end{equation}
where $\boldsymbol{\theta}_u$ is the characteristic temperature of bin $u$, as
defined by \citet{2001ApJ...551..807D}.

The computational cost of evaluating the transition matrix $\mathbf{T}_{ji}$ is
dominated by evaluating the upward transitions, in particular the integrand in
Equation~(\ref{eq:upwardtrans}), separately for each $\{i,j\}$ pair.  It may be
reduced by noting that, although $\mathbf{G}_{ji}$ is a function of $i$ and
$j$, it always remains a linear function of E.  Hence the integrand in
Equation~(\ref{eq:upwardtrans}) is, for any $\{i,j\}$ pair, a linear
combination of:
\begin{equation} 
  I_{1}(a,b)=\int_{a}^{b} dE C_{\rm abs}(E)u_{E} (E)
\end{equation} and
\begin{equation}
  I_{2}(a,b)=\int_{a}^{b} dE \cdot E\cdot C_{\rm abs}(E) u_{E} (E).
\end{equation}
where $I_1$ and $I_2$ are independent of $i$ and $j$. We can then write
$\mathbf{T}_{ji}$ as a linear combination of $I_1(W_1,W_2)$, $I_1(W_2,W_3)$,
$I_1(W_3,W_4)$, $I_2(W_1,W_2)$, $I_2(W_2,W_3)$ and $I_2(W_3,W_4)$, with
coefficients which are independent of $E$.  We evaluate $I_{1}$ and $I_{2}$ by
non-adaptive integration using a pre-computed grid of the two integrands
appearing in $I_{1}$ and $I_{2}$. By using this grid, we reduce the
number of necessary evaluations of $C_{\rm abs}(E)$ and $u_{E} (E)$ from
$\mathcal{O}(N^{2})$ to $\mathcal{O}(10^3)$. Non-adaptive integration is
appropriate since $u_{E}$, the output of stellar population models, is computed
on a fixed grid.
}

\bibliographystyle{mn2e} 
\bibliography{rnote06}

\label{lastpage}
\end{document}